\documentclass[conference]{sty/IEEEtran}
\IEEEoverridecommandlockouts
\usepackage{natbib}
\usepackage{tikz}
\usepackage{amsmath}
\usepackage{xcolor}
\definecolor{autorefcolor}{RGB}{30, 100, 160} 
\usepackage[
  colorlinks=true,
  linkcolor=autorefcolor,     
  citecolor=green!50!black,   
  urlcolor=blue!70!black      
]{hyperref}

\usepackage{amssymb}
\usepackage{filecontents}

\usepackage{xurl}                

\usepackage{amsmath,amsopn}
\usepackage{subfigure}
\usepackage{endnotes,microtype,xspace,graphicx,fancyvrb,multirow}
\usepackage{booktabs}
\usepackage{array,underscore,relsize}
\usepackage[T1]{fontenc}
\usepackage{fancyhdr}
\usepackage{enumitem}
\usepackage[labelfont=bf,font=small,skip=5pt]{caption}
\usepackage{makecell}

\usepackage{fp}
\usepackage{siunitx}

\usepackage{balance}

\usepackage{tikz}
\usetikzlibrary{shapes.geometric,shapes.arrows,decorations.pathmorphing}
\usetikzlibrary{matrix,chains,scopes,positioning,arrows,fit}

\usepackage{algorithm}
\usepackage[noend]{algpseudocode}

\newcommand{\cc}[1]{\mbox{\smaller[0.5]\texttt{#1}}}

\fvset{fontsize=\scriptsize,xleftmargin=8pt,numbers=left,numbersep=5pt}

\input{code/fmt}

\renewcommand{\figurename}{Fig.}

\newif\ifdraft\drafttrue
\newif\ifnotes\notestrue
\ifdraft\else\notesfalse\fi

\input{glyphtounicode}
\newcolumntype{R}[1]{>{\raggedleft\let\newline\\\arraybackslash\hspace{0pt}}p{#1}}

\newcommand{\squishlist}{
\begin{itemize}[noitemsep,nolistsep]
  \setlength{\itemsep}{-0pt}
}
\newcommand{\squishend}{
  \end{itemize}
}

\usepackage{tikz}

\usepackage{xstring}
\newcommand{\PP}[1]{
\vspace{2px}
\noindent{\bf \IfEndWith{#1}{.}{#1}{#1.}}
}

\newcommand{\boxbeg}{
\vspace{2px}
\noindent\begin{tabular}{|l|}\hline
\begin{minipage}{3.2in}
\vspace{2px}
\noindent
}

\newcommand{\boxend}{
\vspace{2px}
\end{minipage}\\ \hline
\end{tabular}
\vspace{-10pt}
}

\newcommand{\ie}{\emph{i.e.,}~}
\newcommand{\eg}{\emph{e.g.,}~}

\usepackage{pifont}

\newcommand*{\unsafe}{\texttt{unsafe}\xspace}

\usepackage{amsmath} 
\usepackage{bm} 
\usepackage{bbm} 
\usepackage{amsthm}
\usepackage{mathtools}

\newcommand{\ep}{\varepsilon}

\makeatletter
\@ifundefined{proposition}{%
    
}{}
\@ifundefined{definition}{%
}{}
\@ifundefined{lemma}{%
    
}{}
\@ifundefined{corollary}{%
    \newtheorem{corollary}{Corollary}
}{}
\@ifundefined{theorem}{%
    
}{}
\@ifundefined{corollary}{%
    
}{}
\@ifundefined{assumption}{%
    
}{}
\@ifundefined{problem}{%
    
}{}
\@ifundefined{problem*}{%
    \newtheorem*{problem*}{Problem}
}{}
\@ifundefined{claim}{%
    
}{}
\@ifundefined{example}{%
    
}{}
\makeatother

\newtheorem{innercustomgeneric}{\customgenericname}
\providecommand{\customgenericname}{}
\newcommand{\newcustomtheorem}[2]{%
  \newenvironment{#1}[1]
  {%
   \renewcommand\customgenericname{#2}%
   \renewcommand\theinnercustomgeneric{##1}%
   \innercustomgeneric
  }
  {\endinnercustomgeneric}
}
\newcustomtheorem{customclaim}{Claim}

\providecommand{\realnum}					{\mathbb{R}}

\renewcommand{\(}						{\left(}
\renewcommand{\)}						{\right)}
\renewcommand{\[}						{\left[}
\renewcommand{\]}						{\right]}

\providecommand{\Prob}{\mathbbm{P}}

\def\sh{\hat{{s}}}

\def\uh{\hat{{u}}}

\def\Us{\mathcal{{U}}}

\def\Xs{\mathcal{{X}}}

\usepackage{fontawesome5}
\usepackage{listings}
\usepackage{utils/nasm/lang}  
\usepackage{utils/nasm/style} 
\usepackage{listings, utils/listings-rust}
\definecolor{backcolour}{rgb}{0.95,0.95,0.92}

\usepackage{booktabs}
\usepackage{bm}
\usepackage[normalem]{ulem}
\usepackage{pifont}
\usepackage{cleveref}
\usepackage{sankey}
\usepackage{xcolor,colortbl}
\usepackage{pgfplotstable}
\usepackage{amsthm}
\usepackage{pgfplots}
\usepackage{tikz}

\usetikzlibrary{shapes,backgrounds,arrows,fit,calc}
\newcommand{\implname}{\mbox{\textsc{Ruby}}}

\providecommand{\unsafe}{unsafe}

\providecommand{\train}{\text{train}}
\providecommand{\val}{\text{val}}
\providecommand{\cali}{\text{cal}}
\providecommand{\test}{\text{test}}

\providecommand{\unsafecls}{unsafe classifier}

\providecommand{\crateunsafe}{CrateU\xspace}
\providecommand{\rustsecunsafe}{RustSecU\xspace}
\providecommand{\rustsecbug}{RustSecB\xspace}

\newcommand{\sqlist}{
	\begin{list}{$\bullet$}
		{ \setlength{\itemsep}{0pt}      \setlength{\parsep}{3pt}
			\setlength{\topsep}{3pt}       \setlength{\partopsep}{0pt}
			\setlength{\leftmargin}{1.0em} \setlength{\labelwidth}{1em}
			\setlength{\labelsep}{0.5em} } }
\newcommand{\sqend}{
	\end{list}  }

\providecommand{\para}[1]{\smallskip\noindent\textbf{#1}}
\providecommand{\finalrecall}{$91.75\%$}
\providecommand{\finalcoverage}{$7.43\%$}

\makeatletter
\newtheorem*{rep@theorem}{\rep@title}
\newcommand{\newreptheorem}[2]{%
\newenvironment{rep#1}[1]{%
 \def\rep@title{#2 \ref{##1}}%
 \begin{rep@theorem}}%
 {\end{rep@theorem}}}
\makeatother

\newreptheorem{theorem}{Theorem}

\begin{document}

\title{Ruby: Unmasking Unsafe Rust in Stripped Binaries via Machine Learning}






\author{\IEEEauthorblockN{Xiang Cheng\IEEEauthorrefmark{1},
Sangdon Park\IEEEauthorrefmark{1}\IEEEauthorrefmark{2}, HyungSeok Han\IEEEauthorrefmark{3},
Xiaokuan Zhang\IEEEauthorrefmark{4} and
Taesoo Kim}
\IEEEauthorblockA{Georgia Institute of Technology,
\IEEEauthorrefmark{2}Pohang University of Science and Technology,
\IEEEauthorrefmark{3}Microsoft,
\IEEEauthorrefmark{4}George Mason University}
}
\maketitle
\def\thefootnote{*}\footnotetext{These authors contributed equally to this work.}

\begin{abstract}
Rust, as an emerging system programming language, introduces \unsafe to allow developers to bypass safety checks during compilation. 
As a result, memory safety bugs are typically confined to the \unsafe regions, which have been the primary focus of Rust bug-finding tools. However, such tools rely on the presence of the \unsafe keyword in Rust source code; there are no tools available that can examine Rust binaries to pinpoint \unsafe areas. Therefore, we propose \implname{}, the first tool that unmasks \unsafe regions in Rust binaries using machine learning. By capturing the subtle differences in the binary instructions, \implname{} can identify 91.75\% of the total \unsafe regions with a false positive rate of 6.16\%, beating SOTA LLM models including GPT-5.2, Claude-4.5 and Gemini-3. We further applied \implname{} to guide symbolic execution and fuzzing, showing a speed-up of 57.95\% and 21.26\%, with five bugs confirmed and patched by Google in Android library fuzzing.

\end{abstract}


\section{Introduction}



Rust is a rapidly growing systems programming language due to its focus on enhancing memory safety~\cite{rust2010} and addressing memory safety vulnerabilities~\cite{msmemorysafetybug, googlememorysafetybug}.
For example, Rust has been employed in system software such as Mozilla Firefox, Google Chrome~\cite{google-security-blog-rust}, Linux kernel modules~\cite{rust-for-linux}, Windows kernel components~\cite{register-microsoft-rust}, and various other device drivers~\cite{rust-gpu-cuda}.

To achieve memory safety while providing flexibility, Rust is divided into safe and unsafe code regions.
Safe Rust is a strongly typed language that ensures memory safety through compile-time checks.
However, its strict rules can limit the ability to implement certain features (e.g., resource sharing, low-level assembly) commonly needed in systems programming.
To overcome this, unsafe Rust is employed, shifting the responsibility of memory safety checks from the compiler to the developers.
In particular, unsafe Rust enables the execution of hazardous operations~\citep{RustUnsafeOps} such as dereferencing pointers or interfacing with external C libraries.

\smallskip\noindent\textbf{Rust Memory Safety Bugs: Source Code Analysis.}
Despite the memory safety mechanism in Rust, more than 360 memory safety bugs have been found in Rust programs in the last five years~\cite{rustsec}.
The primary reason is that the strict rules of safe Rust often necessitate developers to engage with unsafe Rust, which can lead to memory safety bugs when developers fail to manually verify unsafe regions~\cite{astrauskas2020programmers, evans2020rust}.
For example, cyclic types (e.g., doubly-linked lists) cannot be implemented without resorting to unsafe Rust.
Additionally, developers might inadvertently introduce unsafe Rust code when utilizing libraries that contain unsafe regions.

To find and fix memory-safe bugs in Rust programs, a rich line of prior work (\eg~\cite{bae:rudra, xu2024rpg, min2024erasan, li2021mirchecker, cheng:rug, liu2020securing, jiang2021rulf, chen2025typepulse, zhang2025rumono, cui2023safedrop}) has focused on analyzing the unsafe regions in Rust source code.
For example, Rudra~\cite{bae:rudra} presented three important patterns of memory safety bugs in unsafe Rust and identified these bugs through static analysis of the unsafe regions;
RPG~\cite{xu2024rpg} prioritized fuzzing unsafe regions in Rust libraries to more efficiently identify memory safety bugs;
ERASAN~\cite{min2024erasan} proposed a more efficient address sanitizer~\cite{serebryany2012addresssanitizer} for Rust programs, leveraging the characteristics of unsafe regions.

\smallskip\noindent\textbf{Our Focus: Rust Binary Analysis.}
While source-level analysis has advanced significantly in detecting memory safety issues, it is fundamentally constrained by the requirement for source code accessibility. This limitation is increasingly acute as Rust is adopted in commercial and security-critical sectors where proprietary codebases are the norm. For example, when auditing proprietary Rust-based firmware, safety-critical embedded systems, or commercial drivers such as Ferrocene~\cite{ferrocene} and Windows kernel components~\cite{register-microsoft-rust}, security analysts must rely exclusively on the binary. 

Furthermore, binary analysis remains essential even when source code is available, as it provides the ultimate ground truth of the executable's behavior. It can unmask vulnerabilities introduced during the compilation pipeline, such as aggressive compiler optimizations that alter memory access patterns~\cite{wang2012undefined} or complex macro expansions that generate unforeseen \textbf{direct unsafe operations}~\cite{obfstrissue60}. These instruction-level risks are often elided or obscured at the source level, making their detection in stripped binaries a vital necessity for robust security assurance.

For Rust binary analysis, pinpointing unsafe regions is critical since these regions bypass the Rust compiler's memory safety checks and are susceptible to vulnerabilities. 
However, as the \cc{unsafe} keyword is absent from Rust binaries after compilation, identifying these unsafe regions becomes challenging for the following  reasons:

\textbf{$\bullet$ Diverse unsafe Rust operations:} In practice, the rustc compiler checks 12 unsafe operations~\cite{astrauskas2020programmers} outlined in
\autoref{tab:unsafetype}. Each operation presents a distinct root cause, and addressing each unsafe operation constitutes a subproblem of the broader challenge of recovering unsafe Rust from binaries.

\textbf{$\bullet$ Ambiguity between safe/unsafe Rust:} Some unsafe operations can closely resemble safe operations, making them challenging to identify.
Examples include dereferencing pointers or references, traversing unions or structs, and accessing mutable or immutable static variables. 
These operations exhibit minimal differences in the compiled binary; \eg in \autoref{list:deref}, there exists merely a one-instruction difference between the two functions.  

\textbf{$\bullet$ Diverse architectures and compiler toolchains:} Unlike source code, binary programs are closely bound to the target architectures. Rust relies on LLVM as its backend to generate machine code, which does not preserve unsafe information. Therefore, different architectures and compiler toolchains can produce distinct binary programs from the same codebase.

\begin{figure}[t!]
\begin{lstlisting}[language=Rust,  style=boxed, basicstyle=\small\ttfamily, captionpos=b,
caption={Dereferencing a pointer and reference: reference can be optimized out but pointer is compiled into an explicit access. Code are omitted for simplicity and compiled assemblies are in \autoref{list:deref-x64}.}, 
label={list:deref}]
// a is &i32, dereferencing is safe
*a;         // => ; deref op is optimized out
// a is *i32, dereferencing is unsafe
unsafe {*a} // => mov eax,DWORD PTR [rbx]
\end{lstlisting}
\end{figure}

\begin{figure}[]
  \centering

    \centering
    \includegraphics[width=\columnwidth]{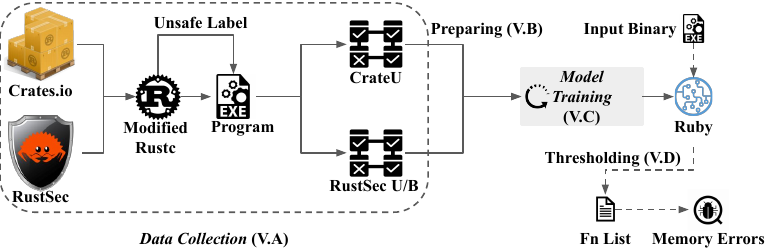}
    
  \caption{
    The workflow of \implname{} as an unsafe Rust classifier, details are shown in \autoref{s:impl}
  }\label{fig:workflow}
\end{figure}

\noindent\textbf{Our Approach.}
Although the diverse unsafe operations and subtle differences from safe operations make it challenging to unmask unsafe Rust, our \textbf{insight} is: there are binary representation differences (see \autoref{sec:design}) between safe and unsafe Rust due to compiler optimizations, because safe Rust provides more constraints.
These subtle differences can be learned by machine learning models during the large scale training phase~\cite{beckman2020binary, shin2015recognizing} and captured through inference. We propose \implname{}, which leverages ML to address the challenges mentioned above
and identify unsafe regions within Rust binaries.

As the workflow of \implname{} shown in \autoref{fig:workflow}, we first construct \crateunsafe{} dataset, which contains
the unsafe labels on functions from the entire \href{https://crates.io}{crates.io}. 
%
Next, \implname{} collects the \rustsecunsafe{} and \rustsecbug{} datasets containing real-world five-year memory safety bugs from Rustsec, with each bug's root cause manually labeled to evaluate \implname{}'s performance.
%
%
We trained \implname{} on \crateunsafe{} and evaluated it on \rustsecunsafe{} and \rustsecbug{}. With the help of the classification confidence score, \implname{} outputs a prioritized list as targets, which can be used by reverse engineers to focus on only 7.43\% of programs, with an 88.92\% chance of finding potential memory safety bugs. In comparison, a default analysis requires examining 41.38\% of the binaries, demonstrating a slowdown of more than 5.5$\times$. We also conducted an evaluation of \implname{} against SOTA LLMs including GPT-5.2, Claude-4.5, and Gemini-3; \implname{} achieves higher accuracy than these general-purpose models.

To illustrate the real-world application of \implname{},
we incorporate \implname{} into the program analysis workflow, using its results as guidance for static analysis by Angr~\cite{shoshitaishvili2016state} and directed fuzzing by AFLGo~\cite{bohme2017directed}.
Utilizing \implname{}, we can achieve a time reduction of 57.95\% in symbolic execution by Angr and 21.26\% in directed fuzzing efforts to identify bugs. By applying \implname{}'s results with Android patch testing for Rust libraries, we successfully identified five bugs in Android's Rust libraries,
which have been confirmed and patched by Google.

In summary, our \textbf{contributions} are as follows:
\sqlist

    \item 
        We propose a new tool \implname{}\footnote{Our artifact is available at \href{https://doi.org/10.5281/zenodo.14217066}{https://doi.org/10.5281/zenodo.14217066}}, which leverages machine learning techniques to address diverse challenges and identify unsafe regions in Rust binaries. To our knowledge, \implname{} is the first tool specifically designed to locate unsafe regions in Rust binaries.  
    \item 
        We extensively evaluate \implname{}, demonstrating that \implname{} is \textbf{effective} (achieving high accuracy), \textbf{efficient} (capable of analyzing \href{https://crates.io}{crates.io} within a week) and \textbf{robust} (compatible with x64 and ARM, and different Rustc/LLVM versions).  
    \item 
        We use \implname{} to identify five bugs in Android Rust libraries, which have been confirmed and patched by Google.
\sqend

\PP{Ethics Discussion.}We collect Rust packages from  \href{https://crates.io}{crates.io}
and bug information from RustSec reports \citep{rustsec}.
All data collected is publicly available, and we only use this data for research purposes. 
For the five bugs we identified in Android, we reported them to Google, and all bugs have been patched. 

\section{Background}
\label{bg}
\begin{table*}
  \centering
  \rowcolors{2}{gray!25}{white}
    \begin{tabular}{c|c|p{75mm}|c}
    \toprule
    Label & Unsafe Operations & \makecell{Description} & \% of functions
        \\
        \midrule
        \makecell{1 \\ 2} & \makecell{\texttt{CallToUnsafeFunction}} &
        \makecell[l]{
          Call an unsafe function. This type has two subtypes: \\
          ``internal'' (label 1) and ``external'' (label 2)
        } & \makecell{17.61\% \\ 0.17\%}
        \\
        3 & \texttt{UseOfInlineAssembly} & Use a \texttt{asm!} macro with low-level assembly.  & 0.01\%
        \\
        \makecell{4} & \makecell{\texttt{InitializingTypeWith}} 
        &
        \makecell[l]{
          Initialize a layout restricted type's field with a value \\
          outside the valid range.
        } & 0.56\%
        \\
        5 & \texttt{CastPointerToInteger} & Cast pointers to integers in const functions. (Deprecated) & 0
        \\
        6 & \texttt{UseOfMutableStatic} & Access to a mutable static variable. & 0.08\%
        \\
        7 & \texttt{UseOfExternStatic} & Access to a mutable static variable from external. & 0.01\%
        \\
        8 & \texttt{DerefOfRawPointer} & Dereference a raw pointer. & 2.59\%
        \\
        9 & \texttt{AccessToUnionField} & Access to a union field. & 1.69\%
        \\
        10 & \texttt{MutationOfLayoutConstrainedField}  & Change the layout of a constrained field. (Deprecated) & 0
        \\
        11 & \texttt{BorrowOfLayoutConstrainedField} & Borrow a layout constrained field. (Deprecated) & 0
        \\
        12 & \texttt{CallToFunctionWith} & Call to a function that requires special target features. & 1.02\%
    \end{tabular}
    \caption{Twelve unsafe operations named internally in Rustc~\cite{rustcunsafety2023} and their distributions (safe Rust takes the 76.28\%~\cite{astrauskas2020programmers}). \implname{} studies their unique binary representations in \autoref{sec:design}.}
    \label{tab:unsafetype}
    \end{table*}

\subsection{Rust Programming Language}
\label{bg:rust}
\textbf{Safe Rust.}
Rust language \citep{matsakis2014rust} provides a memory-safety guaranty at compile time while allowing control over low-level access to resources. The claimed memory-safety guaranty is demonstrated in \citep{Reed2015PatinaA,jung2017rustbelt} under certain assumptions regarding language models. Specifically, safe Rust is defined as: Safe Rust will never cause undefined behavior~\cite{UndefinedBehaviorWiki}.

To achieve memory safety:
safe Rust leverages \emph{ownership} and \emph{borrowing} to ensure that there are no undefined behaviors in the compiled programs \citep{Reed2015PatinaA}. 
Ownership is a relationship between a value and a variable. Each value in Rust is owned by \emph{only one} variable, and the memory associated with the value is automatically freed if the variable goes out of scope. This simple memory management mechanism provides compile-time memory safety without the need for a run-time garbage collector. 
In particular, since a value's associated memory is freed only via \texttt{drop()},
double-free or use-after-free bugs are avoided through the compiler's lifetime checks. 
Although ownership provides a vital safeguard to avoid memory-safety bugs,
it is too strict to allow only one owner for each value.
Thus, borrowing is introduced to address this issue.

Borrowing allows us to reference a variable without
ownership. Specifically, Rust provides two types of borrowing: immutable and
mutable. Each variable can have either multiple immutable references or
only one mutable reference. Through this restriction, safe Rust ensures that no
other party can write to the variable when it is borrowed as a
mutable reference.


\smallskip
\noindent\textbf{Unsafe Rust.} Although safe Rust provides a strong guaranty of memory safety and relatively flexible restrictions,
there are many scenarios where programmers need to maintain a shared mutable reference in system programming.
For example, memory can be shared among multiple threads with well-defined synchronization.  

\begin{figure}[]
\begin{lstlisting}[language=Rust,  style=boxed, basicstyle=\small\ttfamily, captionpos=b,
caption={Example of direct unsafe operations using Rust standard library's \texttt{Vec}. The first two functions are \textit{indirect} unsafe operations, and the third function \cc{get()} is the \textit{direct} unsafe operation.}, 
label={list:direct-unsafe}]
impl<T> Vec<T> {
    ...
    pub unsafe fn set_len(&mut self, 
        new_len: usize) {
        self.len = new_len;
    }
    pub const unsafe fn get_unchecked<I>(&self,
        index: I) -> &I::Output
    {
        unsafe { &*index.get_unchecked(self) }
    }
}
fn get(self, slice: &[T]) -> Option<&T> {
    if self < slice.len() {
        unsafe{Some(slice_get_unchecked(slice,
            self))}
    }
    ...
}
\end{lstlisting}
\end{figure}

To support such cases, unsafe Rust is
introduced to escape from the Rust compiler's checks inside safe regions and
requires programmers to ensure memory 
safety inside the unsafe regions. 
In particular, the Rust compiler defines five operations
as unsafe operations,
which help programmers identify unsafe regions and ensure memory safety~\citep{RustUnsafeOps}: dereference a raw pointer, call an unsafe function or method, access or modify a mutable static variable, implement an unsafe trait, and access fields of unions.
As these memory-related operations within unsafe regions are not checked by
a compiler for memory safety, 
they are likely to cause memory safety bugs.
In practice, the Rust compiler (rustc 1.67) performs explicit checks for 12 distinct unsafe operations~\cite{rustcunsafety2023},
as summarized with their relative frequencies in \autoref{tab:unsafetype}. These 12 explicit operations are derived from the five canonical categories of unsafe operations, and several of them (5, 10, 11) are already deprecated.
Among the remaining nine categories of active unsafe operations, we designate those that do not involve (1) and (2) the invocation of unsafe functions as \textbf{direct unsafe operations}, because \uline{direct unsafe operations are, by themselves, capable of inducing undefined behavior}.

Because Rust’s safety guarantees are defined at the module level~\cite{rustnomiconunsafe}, certain functions, such as \texttt{set\_len}, are declared unsafe yet only cause undefined behavior indirectly, via other direct unsafe operations. An illustrative example is given in \autoref{list:direct-unsafe}: the two unsafe functions shown there do not themselves contain any direct unsafe operations in their bodies but instead rely on the direct unsafe operation in the third function, \texttt{get}, to potentially violate memory safety. In this work, \implname{} primarily targets these direct unsafe operations as exemplified in \texttt{get}.

The direct unsafe Rust takes 24.6\% of the codebase across all crates in the Rust community~\cite{astrauskas2020programmers}, and even if the target project does not have any unsafe code, the usage of third-party and the standard libraries can introduce implicit unsafe regions.

\subsection{Memory Safety Error}
\label{sec:bg-memory}



Memory safety is the state of being protected from various software
bugs and security vulnerabilities when dealing with memory access,
such as buffer overflows and dangling pointers. 
Operations that
violate this protection are considered memory safety errors~\cite{MemorySafetyWiki}. 
Among these errors, access errors, uninitialized variables, and use-after-free can directly enable malicious users to hijack the program, leading to serious consequences.
Therefore, searching for memory safety errors is important for reverse engineers to ensure the program's safety.

On the other hand, these memory safety errors (excluding the memory leaks) are considered undefined behaviors in Rust~\cite{UndefinedBehaviorWiki}, which are designed to be encompassed by unsafe Rust. Ideally, because of the safe Rust definition, undefined behavior (operations) is considered to be triggered only through direct unsafe operations. Therefore, by helping to reduce the search space for reverse engineers in locating direct unsafe operations, \implname{} can accelerate the memory safety error hunting process.

\section{Threat Model}
\label{sec:threat}

We assume the possession of a Rust binary compiled
with the release profile~\cite{RustCargoReleaseProfiles}, without debugging symbols. Furthermore, we are aware of all the function boundaries, which can be analyzed using
reverse engineering tools such as IDA or Ghidra. Besides function boundaries, we do not require any other knowledge about the binary. 

\implname{}'s
objective is to identify the code operations in the binary that are written using
unsafe Rust and may directly trigger undefined behavior, referred to as \textbf{direct unsafe operations}. \implname{} outputs such direct unsafe operations as a sub-region of the whole program operations to minimize manual effort in exploitation. Once these direct unsafe regions
are identified, we can focus on those areas and employ additional intensive analysis (e.g., symbolic execution, directed fuzzing)
to uncover potential vulnerabilities (\autoref{sec:eval:RQ4}).

\section{Binary Representations of Unsafe Rust}
\label{sec:design}


We first study each of the unsafe operations in detail and show the unique binary representations\footnote{All the code samples are compiled and discussed with \texttt{release} profile with Rustc 1.67.0 and linked to executable as output.}.

\subsection{Calling Unsafe Functions}
\label{ssec:callunsafe}

The first category of unsafe operations in Rust is calling unsafe functions. Based on the owner of the unsafe functions, there are two cases for calling unsafe functions: calling unsafe FFI (Foreign Function Interface) functions or calling unsafe Rust functions. An unsafe function indicates that there is an unsafe operation inside the function body, and by adding strict constraints, developers can convert an unsafe function into a safe function. 
\subsubsection{Calling unsafe Rust Function}
Calling unsafe Rust functions (\ie label 1) indicates that both the caller and callee are from Rust. It is a language-level definition of unsafe because calling an unsafe Rust function does not directly trigger undefined behaviors or memory safety errors, but it is the unsafe operations inside the unsafe functions that trigger these bugs.

To correctly infer such functions, there are two challenges:

\sqlist
\item By adding certain restrictions, developers can wrap an unsafe function into a safe function. However, to detect unsafe Rust functions, \implname{} is expected to reason these complex restrictions to check if they covered all cases, which is as difficult as the NP hard problem~\cite{king1976symbolic}.
\item To audit the target function, all callee functions and their inner functions are expected to be added to the context, exceeding the token limitation for machine learning models. Therefore, directly detecting unsafe Rust is challenging both for static analysis and machine learning models.
\sqend

Instead of directly identifying such cases, \implname{} tries to identify the root unsafe operations inside the unsafe function and skip the unsafe calls. 
\implname{}'s solution is based on an observation under \implname{}'s threat model:

\textbf{Direct Unsafe Delegation:} Every unsafe Rust function must eventually ground its unsafety in a direct unsafe operation other than the call itself.

We separate this property into two individual problems: 

\smallskip\noindent\textbf{Problem 1.} The Rust unsafe calling function must have a different innermost unsafe operation.

Suppose we have an unsafe function $F$ in Rust without root cause operation, which means that the Rustc compiler cannot detect any unsafe operations inside $F$. Then we can safely remove the \texttt{unsafe} keyword from $F$ without causing compilation errors. Thus, for any unsafe Rust functions, there must be an inner unsafe operation inside the function body, referring to its root cause.

\smallskip\noindent\textbf{Problem 2.} The innermost unsafe operation within the unsafe function is not calling to an unsafe Rust function.

Suppose that we have a function $F_{n+1}$ calling an unsafe function $F_{n}$, and the caller $F_{n+1}$ will have unsafe regions around the calling instructions. For the callee function $F_{n}$, its unsafe label can be either because it calls other unsafe Rust functions (\ie label 1) or it has other unsafe operations (\ie label 2-12) in \autoref{tab:unsafetype}. For the second case, we show that the root cause is different from the callee itself. For the first case, there must exist another unsafe Rust function $F_{n-1}$ called $F_n$, so the root cause of $F_n$ will be delegated to $F_{n-1}$. Due to \texttt{Problem 1}'s proof, there must exist an innermost function $F_{0}$ that could not be further delegated, and the unsafe operation $F_0$ belongs to the second case.

Guided by this observation, \implname{} intentionally skips the identification of the \texttt{call} instruction itself. Instead, it attempts to locate the concrete direct unsafe operations that serve as the root cause of the unsafety. By pinpointing these "root" locations, downstream tools like directed fuzzing can more effectively target the specific instructions where memory corruption is physically possible, rather than wasting resources on high-level function entries.

\subsubsection{Calling unsafe FFI Function} FFI functions enable Rust to interact with existing C/C++ libraries and are inherently unsafe. These functions can originate from either statically or dynamically linked libraries. For dynamically linked functions, \implname{} embeds the dynamic symbol table entries corresponding to the called functions as part of the input, aiding the model in recognizing such cases. In the case of statically linked functions, Rust functions often exhibit different stack prologues from their C/C++ counterparts. Due to ownership tracking and move semantics, Rust tends to allocate more local variables, resulting in larger stack frames compared to typical C/C++ functions. An example can be found at \href{https://godbolt.org/z/4G5cYoroK}{Godbolt}. These differences can be captured by ML models to distinguish the function sources. 




\subsection{Using Inline Assembly}

\begin{figure}[]
\begin{lstlisting}[language=Rust, style=boxed, basicstyle=\small\ttfamily, captionpos=b,
caption=Example of using inline assembly in Rust. The hand-written assembly is preserved in the final binary whereas equivalent Rust code is optimized by compiler. Full example can be found at \href{https://godbolt.org/z/e1eM5saq9}{Godbolt},
label={list:use-asm}]
let mut x: u64 = 4;
asm!(
    "mov {tmp}, {x}",
    "shl {tmp}, 1",
    "shl {x}, 2",
    "add {x}, {tmp}",
    x = inout(reg) x,
    tmp = out(reg) _,
);           // Assembly preserved in binary
let mut y: u64 = 4;
let tmp = y << 1;
y = y << 2;
y = y + tmp; // Compiler optimizes computation
\end{lstlisting}
\end{figure}

Rust, similar to C/C++, enables developers to integrate inline assembly for precise control over low-level behavior. These inline instructions are treated as \texttt{unsafe} since they bypass the Rust compiler’s type and memory safety checks. As shown in \autoref{list:use-asm}, developer-written assembly is preserved verbatim in the final binary, whereas semantically equivalent Rust code is typically optimized away or transformed. Moreover, manually crafted assembly usually varies from compiler-generated code, and these discrepancies can be utilized to detect potential unsafe operations during binary analysis.
However, we note that developers may write arbitrary inline assembly due to its flexibility, creating challenges for \implname{} to identify.

\subsection{Initializing Type with Constraints}

\begin{figure}[]
\begin{lstlisting}[language=Rust,  style=boxed, basicstyle=\small\ttfamily, captionpos=b,
caption={Example of using \texttt{rustc\_layout\_scalar\_valid\_range} to enable the niche optimizations for Rustc, leading to different memory layout. The full example can be found in \href{https://godbolt.org/z/h67fjj4dd}{Godbolt}}, 
label={list:init-type}]
#[rustc_layout_scalar_valid_range_start(1)]
#[rustc_layout_scalar_valid_range_end(5)]
struct NonZeroI64(i64);
size_of::<Option<NonZeroI64>> // 8
#[repr(transparent)]
struct PlainI64(i64);
size_of::<Option<PlainI64>> // 16
\end{lstlisting}
\end{figure}

To further support compiler optimization, Rust provides \texttt{rustc\_layout\_scalar\_valid\_range} attributes that enable users to specify the valid range of a given variable. Any value outside of the valid range can lead to invalid values, which is considered unsafe in Rust. The compiler can then assume this valid range and perform niche optimizations accordingly. As shown in \autoref{list:init-type}, niche optimization~\cite{atticusnicheoptimizations} in Rust attempts to compress the actual memory size of enums with values by reusing invalid values of the wrapped type. In the example, \texttt{NonZeroI64} is defined with a valid range from 1 to 5, making values outside this range—including 0—invalid. This enables the compiler to use 0 as a niche value to represent \texttt{None}, reducing the size of \texttt{Option<NonZeroI64>} to 8 bytes, the same as i64. In contrast, \texttt{PlainI64} does not provide such a range guarantee, thus \texttt{Option<PlainI64>} cannot reuse any value as a niche and requires 16 bytes to store both the value and the discriminant.
By inspecting the memory access patterns, \implname{} can analyze the memory size of the types and attempt to identify this type of unsafe operation.




\subsection{Mutating Static Variables}

Mutating global variables can lead to data races, which are considered unsafe in Rust. To mitigate this, Rustc explicitly checks two types of unsafe operations: (1) mutations to static variables defined in Rust code, and (2) accesses to static variables originating from FFI libraries (e.g., in C/C++).

To support detection in binary analysis, \implname{} embeds information from static variable sections (i.e., \texttt{.data} and \texttt{.bss} in ELF binaries) into the corresponding functions, enabling the model to identify and reason about global variable usage.

For FFI static variables, Rustc cannot analyze their mutability, as their definitions reside outside of its analysis scope. Consequently, it conservatively assumes all FFI static variables are mutable and treats any access to them as unsafe. As discussed in \autoref{ssec:callunsafe}, \implname{} can leverage characteristic differences between Rust and C/C++ binaries to infer the origins of these static variables and assess their safety.

\subsection{Dereferencing Raw Pointers}

\begin{figure}
  \centering
  \begin{lstlisting}[
      language=nasm,
      style=nasm,
      backgroundcolor=\color{backcolour},
      numbers=left,
      numbersep=5pt,
      captionpos=b,
      tabsize=1,
      caption={Different binary instruction outputs for code shown in \autoref{list:deref} to access pointers/references. The safe guarantee of reference promotes its value to registers. Check full example at \href{https://godbolt.org/z/xzGc4PE49}{Godbolt}.},
      label={list:deref-x64}
    ]
<reference>:
    ... ; omitted 13 instructions for simplicity
    ; deref reference is optimized and removed
    8eb4:	add    rsp,0x40
    8eb8:	pop    rbx
    8eb9:	ret 

<pointer>:
    ... ; omitted 13 instructions for simplicity
    8f33:	mov    eax,DWORD PTR [rbx]; deref ptr
    8f35:	add    rsp,0x40
    8f39:	pop    rbx
    8f3a:	ret
  \end{lstlisting}
\end{figure}



Dereferencing raw pointers, a common operation in C/C++, can lead to severe memory safety issues, such as use-after-free and out-of-bounds access. To uphold memory safety, Rust enforces the use of references instead of raw pointers through its borrow checker, as illustrated in \autoref{list:deref}. 

However, from the perspective of binary instructions, both references and raw pointers are represented as memory addresses, making them indistinguishable at the instruction level. This poses a key challenge in binary analysis: correctly identifying whether a given memory access corresponds to a safe reference or an unsafe raw pointer.

To address this, we observe that Rust references act as `checked pointers' with guaranteed valid access. This property enables the compiler to optimize references more aggressively, frequently promoting them to registers and eliminating redundant memory loads. In contrast, raw pointer accesses lack these guarantees and are generally preserved explicitly in the binary. As demonstrated in \autoref{list:deref-x64}, the compiler optimizes reference-based access into register operations, whereas raw pointer dereferencing produces additional memory access instructions.




\subsection{Accessing Union Fields}

\begin{figure}[]
\begin{lstlisting}[language=Rust, style=boxed, basicstyle=\small\ttfamily, captionpos=b,
caption={Example of accessing union fields in Rust. Each field access generates assembly with different operand sizes, reflecting their types. Full example can be found at \href{https://godbolt.org/z/j4s8nK3PT}{Godbolt}},
label={list:union-access}]
union MyUnion {
    f1: u32,
    f2: i64,
}
let mut x = MyUnion { f1: 1 };
x.f1 = x.f1 + 1; // mov DWORD PTR [rsp],0x2
x.f2 = x.f2 + 1; // inc QWORD PTR [rsp]
\end{lstlisting}
\end{figure}

Unions are special types that allow multiple fields of different types to share the same memory location. While union types are powerful and flexible, improper initialization or access can lead to undefined behavior, such as reading uninitialized memory or interpreting bits with an incorrect type.

At the binary level, union accesses can be distinguished by the operand size of the generated instructions, which reflects the accessed field’s type. As shown in \autoref{list:union-access}, accessing the 32-bit field results in a \texttt{DWORD} instruction, while accessing the 64-bit field results in a \texttt{QWORD} instruction. \implname{} leverages these differences in operand size to infer possible field types in union-based memory locations and to detect potentially unsafe or type-violating operations.




\subsection{Calling Functions with Hardware Features}

\begin{figure}[]
\begin{lstlisting}[language=Rust, style=boxed, basicstyle=\small\ttfamily, captionpos=b,
caption={Example of using AES-NI instructions in unsafe Rust, resulting in \texttt{aesenclast} instructions in assembly. Full example can be found at \href{https://godbolt.org/z/cjx57Kv35}{Godbolt}},
label={list:hardware}]
#[target_feature(enable = "aes")]
...
for &key in &keys[1..KEYS - 1] {
    b = _mm_aesenc_si128(b, key);
} // aesenclast xmm0, XMMWORD PTR [rip+0x3d533]
\end{lstlisting}
\end{figure}

Rust provides the \texttt{std::arch} module to enable direct use of specialized hardware instructions, such as SIMD and AES-NI, for performance-critical operations. However, these instructions depend on specific CPU features and may not be supported across all hardware platforms. To manage compatibility, Rust uses the \texttt{target\_feature} attribute at the function level to indicate the required hardware capabilities. Executing such instructions on unsupported hardware can result in undefined behavior.
In \autoref{list:hardware}, the AES-NI instruction \texttt{_mm\_aesenc\_si128} is utilized to enhance the performance of AES encryption. This leads to the inclusion of the \texttt{aesenclast} instruction within the binary, which is tailored for the x86\_64 architecture. These hardware-dependent instructions are preserved in the binary and can be identified by \implname{}.



\subsection{Phantom Unsafe Operations}

Certain unsafe operations in Rust, such as \texttt{CastPointerToInteger}, \texttt{MutLayoutConstrainedField}, and \texttt{BorrowLayoutConstrainedField}, are considered \textbf{phantom}—they exist in the Rust compiler's internal semantics (e.g., in the \texttt{rustc} implementation) but are virtually absent in real-world Rust code. Our dataset, which includes over 150K crates from \href{https://crates.io}{crates.io}, contains no observed instances of these operations. Due to their rarity and limited practical use in Rust development, \implname{} is not trained or designed to detect these phantom unsafe operations in binary programs and excludes them from its analysis scope.

\section{Proposed Approach}
\label{s:impl}


    

Building on the insights of binary representations in \autoref{sec:design}, we propose \implname{}, a machine learning-based tool for detecting unsafe operations. 
In this section, we present the design choices and implementation details of \implname{}.

\subsection{Dataset Collection}

\begin{table}[]
  \centering
  \footnotesize
  \begin{tabular}{c||c|c}
    \toprule
    Label & \crateunsafe{} & RustSec U/B \\
    \midrule
    \midrule
    safe & $689,105,165$ ($76.28\%$) & $9,001,339$ ($79.16\%$) \\
    unsafe & $214,246,312$ ($23.72\%$)& $2,370,234$ ($20.84\%$) \\
    \midrule
    no-bug & - & $1,815,230$ ($99.98\%$) \\
    bug & - & $302$ ($0.02\%$) \\
    \bottomrule
  \end{tabular}
  \caption{Dataset statistics in the number of functions/records under x86\_64 architecture. The \rustsecunsafe{} contains safe/unsafe labels and \rustsecbug{} contains bug labels.}
  \label{tab:datastats}
\end{table}

\autoref{tab:datastats} summarizes the statistics of the datasets in the number of
functions in binaries along with their labels. 

\PP{Crate}The Crate dataset, denoted by \emph{\crateunsafe{}},
is a set of pairs consisting of a function in binary and the corresponding unsafe labels,
\ie $\{(x_1, u_1), \dots,$
$ (x_m, u_m)\}$,
where
$x_i$ is a function in binary,
$u_i$ is a set of safe or unsafe labels
(\ie $u_i=\{ 0 \}$ for ``safe'' and
$u_i \subseteq \{ 1, ..., 12\}$ for ``unsafe'' root causes in \autoref{tab:unsafetype}),
$m$ is the total number of function and label pairs.
The dataset is generated from all the Rust crates from \href{https://crates.io}{crates.io}, encompassing a total of 107,460 crates.

\PP{RustSec}To further demonstrate the connection between unsafe Rust and memory safety errors, we created a novel dataset
that contains real cases of memory safety bugs from the RustSec Advisory Database \citep{rustsec}. The RustSec dataset is a set of
 tuples of a function in assembly code, unsafe labels, and a bug label, \ie
$\{(x_1, u_1, y_1), \dots, (x_n, u_n, y_n)\}$,
where
$x_i$ is a function,
$u_i$ is a set of unsafe labels as in \crateunsafe{}, 
$y_i$ is a bug label,
and
$n$ is the total number of labeled functions. The combination of labeled functions only with unsafe labels is denoted as the \rustsecunsafe{} dataset, and only with bug labels is denoted as the \rustsecbug{} dataset. During the construction of these datasets, we exclude the relevant crates from \crateunsafe{} to ensure that the model will not be trained with them.

To evaluate \implname{} on real-world vulnerabilities, we collected 360 advisories from the RustSec Advisory Database~\cite{rustsec} spanning five years. Two people independently audited these reports to identify reproducible memory-safety vulnerabilities, filtering the set to 121 unique root-cause bugs by excluding logic-only issues. We downloaded the 257 corresponding crates and compiled them to generate our evaluation dataset, \rustsecbug{}. To prevent data leakage, these 257 crates were strictly excluded from the 11M-function \crateunsafe{} training set. 

While the dataset contains 121 unique vulnerabilities, they manifest as 301 buggy function entries in the compiled stripped binaries. This expansion is a technical consequence of the Rust compilation model: a single generic vulnerability is often specialized into multiple concrete instances via \textit{monomorphization} or propagated across function boundaries through \textit{inlining}. In total, the \rustsecbug{} evaluation set comprises 1.8M functions, of which only 301 are buggy ($<0.02\%$). This extreme class imbalance underscores the "needle-in-a-haystack" challenge of binary-level bug hunting and the necessity of \implname{}'s search-space reduction.

\para{Generation.} To generate the \crateunsafe{} dataset for training purposes, \implname{} utilizes two components: a custom Rust toolchain to record unsafe locations and a binary analyzer to map instructions back to the source. The process begins by modifying the Rustc to log the locations of all unsafe operations during the compilation process. Subsequently, the modified toolchains are applied to recompile the input crates and extract the unsafe location information from both the compiler and binary programs as compilation output. Additionally, the configuration files are modified to include debugging output for all compilation targets. After obtaining the binary programs and compilation logs, binary analyzer can utilize the DWARF debugging information present in the binary programs to map the instructions back to the source code. By comparing the source code locations of unsafe regions, the binary analyzer outputs the unsafe region addresses and labels as part of the training dataset. Through this automated approach, \implname{} builds the \crateunsafe{} dataset for  training and \rustsecunsafe{} evaluation.

\subsection{Preprocessing}


\PP{Embedding Metadata}As discussed in \autoref{sec:design}, detecting certain unsafe operations related to global variables and FFI functions necessitates a comprehensive understanding of the binary program's metadata and memory layout. Thus, \implname{} incorporates the binary's metadata information:
(1) the static analysis of global variables is represented as a special token \texttt{<GLB>} when instructions access the global variables in the \texttt{.data} and \texttt{.bss}, and
(2) external function calls are signified as \texttt{<EXT>} when the calling instruction attempts to invoke an external function. 

\implname{} embeds these two metadata as special tokens in the same line as the assembly instruction and utilizes machine learning models to capture subtle differences in assembly instructions, inferring correct unsafe operations.

\PP{Tokenizing}Before the training step, \implname{} first trains a customized tokenizer based on the input architecture and assembly language. Since assembly is a specialized programming language, a customized tokenizer can effectively split the assembly instructions into meaningful tokens without compromising their integrity. \implname{} by default applies its analysis at the function granularity; however, for functions longer than the token limitation of the model, \implname{} will segment the instructions into pieces that conform to the token limitation and perform analysis on each piece independently. After analyzing all the pieces, \implname{} will consolidate the results to produce the final output.

\PP{Sampling}Due to the large number of records in \crateunsafe{} dataset and limited computing resources, \implname{} we cannot perform training and evaluation on the entire dataset. Consequently, we sampled 10 million records from the entire \crateunsafe{} dataset (approximately 806 million in total) for training purposes. During the sampling process, \implname{} we prioritized the minor labels by attempting to include all cases while maintaining a 1:1 ratio of safe to unsafe records (the biased distribution is preserved in the validation and test datasets). Specifically, based on the training dataset, \implname{} we calculate the weights of each label to support a weighted loss function and mitigate bias during the training process. 
\subsection{Model Training}

We define our problem as  a multi-class, multi-label classification task, where the model input is a sequence of binary instructions in assembly language, and the expected output consists of labels for the input sequences that indicate their unsafe status and reasons. We note from Rust's unsafe definition that the labels can overlap (e.g. accessing a global mutable union structure); hence, the expected output can be either \texttt{safe} label or \texttt{unsafe} label, with at least one reason representing the root cause of the unsafe regions. 

\PP{Model Definition}The goal of unsafe classification is
to design a classifier that predicts
whether a given function embeds unsafe blocks.
In particular, let
$x \in \Xs$ be a sequence of instructions represented in assembly code,

let
$\Us \coloneqq \{2,3,4,6,7,8,9,12\}$ be a set of unsafe labels, where
the corresponding unsafe notations are defined in \autoref{tab:unsafetype},

$u \in 2^{\Us \cup \{0\}}$ be a subset of safe or unsafe labels, where a safe label is denoted by $0$,
$\sh: \Xs \times \Us \cup \{0\} \rightarrow \realnum_{\geq 0}$ be an unsafe scoring function,
and
$\uh: \Xs \rightarrow \{0, 1\}$ be the binary unsafe classifier.

Lastly, labeled functions from a distribution are split into
train, validation, and test sets \ie
$S \coloneqq (S_\train, S_\val, S_\test)$.

We consider the following parameterized function of the binary unsafe classifier based on $\sh$:
\begin{align*}
  \uh(x) \coloneqq
  \begin{cases}
    1 &\text{if~} 1 - \sh(x, 0) \ge \hat\tau \\
    0 &\text{otherwise}
  \end{cases}.
\end{align*}
Here,
$\sh(x, 0)$ is the safe score for a given function $x$ and
$\hat\tau \in \realnum_{\ge 0}$ is a threshold for \unsafecls;
thus, if the risk $1 - \sh(x, 0)$ is greater than the threshold $\hat\tau$,
we consider a function $x$ to be unsafe.

\PP{Model Structure}We use RoBERTa-large \citep{liu2019roberta} as the backbone of the unsafe classifier, 
which is the long-standing stable masked language model based on transformers \citep{vaswani2017attention} and
its input is assembly code. 
On top of this backbone model, we attach a classification header for the unsafe classifier. 
The entire unsafe classifier is trained by minimizing the loss of cross entropy in the training set $S_\train$ for each unsafe and safe label. In addition, we weighted each label's loss by considering the distribution in the training dataset to eliminate the effects of the unfair label distribution.

\subsection{Trustworthy Thresholding}
\label{ssec:pacthresholding}
After the training process, we get the unsafe classifier, noted at $\uh$. Now we describe how to pick the threshold $\hat\tau$
with a guarantee of correctness on
the recall of $\uh$. 
In particular, choosing a threshold for a classifier is
a classic problem \citep{platt1999probabilistic},
where heuristic methods are mostly considered.
We consider a rigorous thresholding approach that comes with a PAC guarantee
based on conformal prediction \citep{vovk2013conditional,wilks1941determination}.


\PP{PAC Algorithm}We adopt the PAC conformal set algorithm \citep{Park2020PAC,park2022pac} for thresholding.
Let $\bar{\theta}$ be the upper Clopper-Pearson (CP) bound \citep{clopper1934use},
where the binomial parameter $\mu$ is included with high probability, \ie
$
  \bar{\theta}(k; m, \delta) \coloneqq
  \inf\{ \theta \in [0, 1] \mid F(k; m, \theta) \le \delta\} \cup \{1\},
$
where $\Prob_{k \sim \text{Binomial}(m, \mu)}\[ \mu \le \bar{\theta}(k; m, \delta)\] \ge 1 - \delta$.
Here, $F(k; m, \theta)$ is the cumulative distribution function of the
binomial distribution with trials $m$ and the probability of success $\theta$.
The threshold $\hat\tau$ is obtained by solving the following optimization:
\begin{align}
  \hat\tau =
  \arg\max_{\tau \in \realnum_{\ge 0}}~ \tau \qquad
  \text{subj. to} \qquad \bar{\theta}(k; |S_\cali|, \delta) \le \ep,
  \label{eq:alg}
\end{align}
where
$S_\cali$ is the set of unsafe functions in $S_\val$, \ie
$S_\cali \coloneqq \{ (x, u) \in S_\val \mid u \neq \{0\} \}$,
and
$k$ is the number of unsafe functions that are missed by a threshold, \ie
$k \coloneqq \sum_{(x, u) \in S_\cali} \mathbbm{1}\( 1 - \sh(x, 0) < \tau \)$.
Intuitively, the interval $[\hat\tau, \infty)$ contains
the most unsafe scores $1 - \sh(x, 0)$ for $x \in S_\cali$.
If an downstream analyzer wants to have $90\%$ recall on unsafe functions,
$\ep$ is set by $0.1$;
if the analyzer wants this desired recall level to be strictly satisfied,
$\delta$ needs to be small, where we use $\delta = 10^{-3}$.





\section{Evaluation}
\label{sec:eval}

To demonstrate the effectiveness of \implname{}, we conducted a comprehensive evaluation motivated by the following questions:

\sqlist
\item \textbf{RQ1:} How effective is \implname{} for unsafe Rust classification?
\item \textbf{RQ2:} How does unsafe Rust help for reverse engineering?
\item \textbf{RQ3:} How robust is \implname{} when dealing with different architectures and different compiler toolchains?
\item \textbf{RQ4:} How effective is \implname{}'s guidance on practical vulnerability hunting process?

\sqend

\para{Hardware Settings.} We launch our experiment on a machine with 128-core AMD EPYC 7452 processors and 8 NVIDIA RTX A6000 GPUs running under the Ubuntu 22.04 operating system. For ARM evaluation, we use an ARM Neoverse-N1.

\subsection{RQ1: Unsafe Classifier Evaluation}\label{exp:unsafeeval}


\begin{figure}[]
  \centering

    \centering
    \includegraphics[width=\columnwidth]{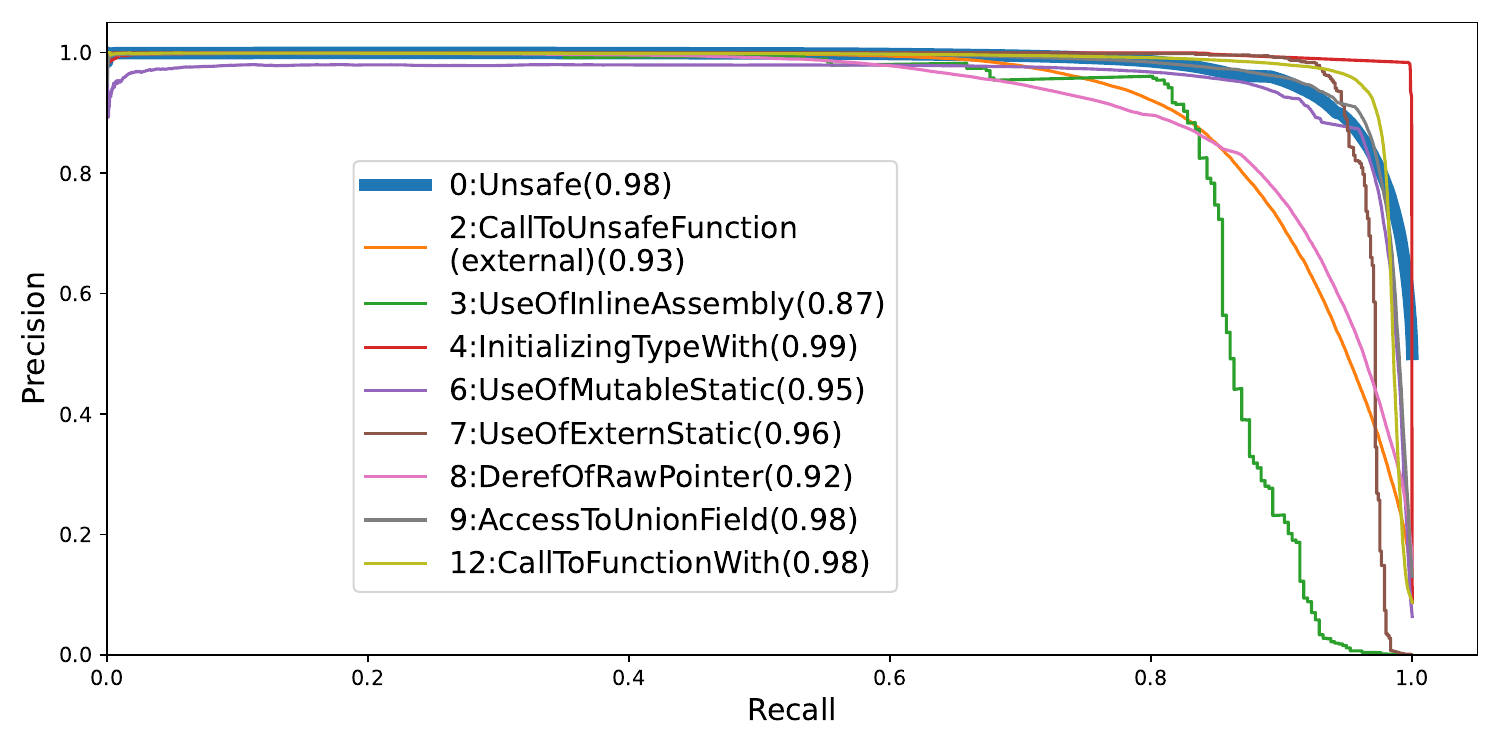}
    \label{fig:unsafeclseval:x64}


  \caption{
    Precision-recall(AUPRC) evaluation on x64 binaries  in \crateunsafe{} dataset. \implname{} achieves relatively high scores on each label and 0.98 on the unsafe detection.
  }
  \label{fig:unsafeclseval}
\end{figure}

\smallskip\noindent\textbf{Setup.} We utilize the \crateunsafe{} dataset for both training and evaluation. The dataset is partitioned into 60\% for training, 20\% for validation during training, and 20\% for testing. We also conduct evaluation on \rustsecunsafe{}, which are excluded from \crateunsafe{} to ensure that \implname{} never learned from those data.


\subsubsection{Evaluation on \crateunsafe{}}
We first evaluate \implname{} over \crateunsafe{} for the unsafe classification task.
\autoref{fig:unsafeclseval} presents the precision-recall curve in \rustsecunsafe{}
for each type of unsafe.

\smallskip\noindent\textbf{Unsafe Classifier.}
The area under the precision recall curve (AUPRC) of
an \unsafecls{} is 0.98 for the overall safe/unsafe classification, demonstrating that
unsafe blocks in the Rust binary are identifiable. Besides, \implname{} applied weights to eliminate the biased distribution of unsafe labels, achieving higher scores in all unsafe Rust classification tasks.


\smallskip\noindent\textbf{Trustworthy Thresholding.}
The precision-recall curve shows the trend of precision and recall with a varying
threshold $\hat\tau$; however, this threshold should be chosen in practice.
We use the trusted thresholding algorithm proposed in (\ref{eq:alg}) for $\hat\tau$, and
the chosen threshold provides $91.75\%$ recall of unsafe functions, which is larger than the desired recall of $90\%$, as expected. 
This suggests that reliable thresholding provides the desired guaranty of recall, controlled by $\ep$.
In practice, we desire to have a list of functions containing the desired rate of unsafe functions, so
we empirically demonstrate that the proposed algorithm achieves this goal.

\smallskip\noindent\textbf{Comparison with SOTA LLMs.} We compare \implname{} model with thresholding to the SOTA LLMs, including chat models: GPT-5.2~\cite{openai2025gpt52}, Claude-Sonnet-4.5~\cite{anthropic2025claude45sonnet}, Gemini-3-flash, and reasoning models: Claude-Opus-4.6~\cite{anthropic2026claude46opus}. We conduct few-shot (K) and best-of-N evaluations to validate the abilities of the SOTA models with gradually increased context. The few-shot evaluation involves gradually adding K examples for each candidate category in the prompt and evaluating the model's inference based on existing examples. For the best-of-N evaluation, we ask the model N times for the same question and check if there is a chance that the model outputs the correct answer. The evaluation results are shown in \autoref{tab:sotallm}.

Overall, \implname{} successfully lead the accuracy by around 20\% as an dedicated model for unsafe Rust classification. Although few-shot and best of N evaluations help SOTA LLMs to better understand the problem, they still suffer from lack of context and requires further improvement.

\smallskip\noindent\textbf{Comparison with Fine-tuned Models.} To further demonstrate \implname{}'s benefit in the specific domain of the Rust binary unsafe classification task, we conduct fine-tuning using OpenAI's GPT-4\footnote{As of February 2026, gpt-4.1-2025-04-14 is the latest model supporting fine-tuning in OpenAI's services.}. For each label, we sampled 20 cases for each unsafe label and combined them into our fine-tuning dataset, reusing the same benchmark that we tested with the vanilla models. The result is shown in the \texttt{GPT-4.1-FT} row in the \autoref{tab:sotallm}. By comparing the result, we believe it's more efficient for \implname{} to adopt a dedicated model instead of general LLM models.

\begin{table}[]
\resizebox{\columnwidth}{!}{%
\begin{tabular}{|c|ccccccccc|}
\hline
\multirow{2}{*}{\begin{tabular}[c]{@{}c@{}}Model\\ Accuracy \% / N=?\end{tabular}} &
  \multicolumn{3}{c|}{K=1} &
  \multicolumn{3}{c|}{K=3} &
  \multicolumn{3}{c|}{K=5} \\ \cline{2-10} 
 &
  \multicolumn{1}{c|}{1} &
  \multicolumn{1}{c|}{3} &
  \multicolumn{1}{c|}{5} &
  \multicolumn{1}{c|}{1} &
  \multicolumn{1}{c|}{3} &
  \multicolumn{1}{c|}{5} &
  \multicolumn{1}{c|}{1} &
  \multicolumn{1}{c|}{3} &
  5 \\ \hline
Sonnet-4.5 &
  \multicolumn{1}{c|}{12.3} &
  \multicolumn{1}{c|}{13.7} &
  \multicolumn{1}{c|}{13.4} &
  \multicolumn{1}{c|}{24.0} &
  \multicolumn{1}{c|}{25.7} &
  \multicolumn{1}{c|}{22.6} &
  \multicolumn{1}{c|}{32.0} &
  \multicolumn{1}{c|}{31.1} &
  30.0 \\ \hline
Gemini3-flash &
  \multicolumn{1}{c|}{38.6} &
  \multicolumn{1}{c|}{41.1} &
  \multicolumn{1}{c|}{41.7} &
  \multicolumn{1}{c|}{61.4} &
  \multicolumn{1}{c|}{62.3} &
  \multicolumn{1}{c|}{62.0} &
  \multicolumn{1}{c|}{63.7} &
  \multicolumn{1}{c|}{64.3} &
  \textbf{65.1} \\ \hline
GPT-5-chat &
  \multicolumn{1}{c|}{43.4} &
  \multicolumn{1}{c|}{48.3} &
  \multicolumn{1}{c|}{48.9} &
  \multicolumn{1}{c|}{55.4} &
  \multicolumn{1}{c|}{57.7} &
  \multicolumn{1}{c|}{57.8} &
  \multicolumn{1}{c|}{58.8} &
  \multicolumn{1}{c|}{62.6} &
  63.4 \\ \hline
Opus-4.6 &
  \multicolumn{1}{c|}{49.7} &
  \multicolumn{1}{c|}{50.3} &
  \multicolumn{1}{c|}{49.4} &
  \multicolumn{1}{c|}{59.7} &
  \multicolumn{1}{c|}{60.3} &
  \multicolumn{1}{c|}{60.9} &
  \multicolumn{1}{c|}{65.4} &
  \multicolumn{1}{c|}{65.7} &
  \textbf{66.1} \\ \hline
Geimin3-pro &
  \multicolumn{1}{c|}{33.1} &
  \multicolumn{1}{c|}{47.1} &
  \multicolumn{1}{c|}{52.5} &
  \multicolumn{1}{c|}{49.7} &
  \multicolumn{1}{c|}{56.6} &
  \multicolumn{1}{c|}{62.6} &
  \multicolumn{1}{c|}{47.4} &
  \multicolumn{1}{c|}{55.7} &
  60.6 \\ \hline
GPT-5.2-pro &
  \multicolumn{1}{c|}{44.0} &
  \multicolumn{1}{c|}{47.7} &
  \multicolumn{1}{c|}{49.1} &
  \multicolumn{1}{c|}{54.0} &
  \multicolumn{1}{c|}{58.3} &
  \multicolumn{1}{c|}{59.7} &
  \multicolumn{1}{c|}{57.7} &
  \multicolumn{1}{c|}{62.3} &
  62.6 \\ \hline
GPT-4.1-FT &
  \multicolumn{1}{c|}{58.0} &
  \multicolumn{1}{c|}{62.3} &
  \multicolumn{1}{c|}{64.0} &
  \multicolumn{1}{c|}{64.0} &
  \multicolumn{1}{c|}{69.1} &
  \multicolumn{1}{c|}{72.3} &
  \multicolumn{1}{c|}{64.3} &
  \multicolumn{1}{c|}{69.2} &
  \textbf{74.6} \\ \hline
\implname{} &
  \multicolumn{9}{c|}{83.1} \\ \hline
\end{tabular}%
}
  \caption{
    Accuracy score of \implname{}'s unsafe classification comparing with SOTA LLM models. We conduct few-shot (K) and best of N evaluations on chat and reasoning models, overall \implname{} improves the accuracy score by training on unsafe classification tasks. 
  }
\label{tab:sotallm}
\end{table}

\subsubsection{Evaluation on \rustsecunsafe{}}
\begin{figure}[]
  \centering
    \centering
    \includegraphics[width=\columnwidth]{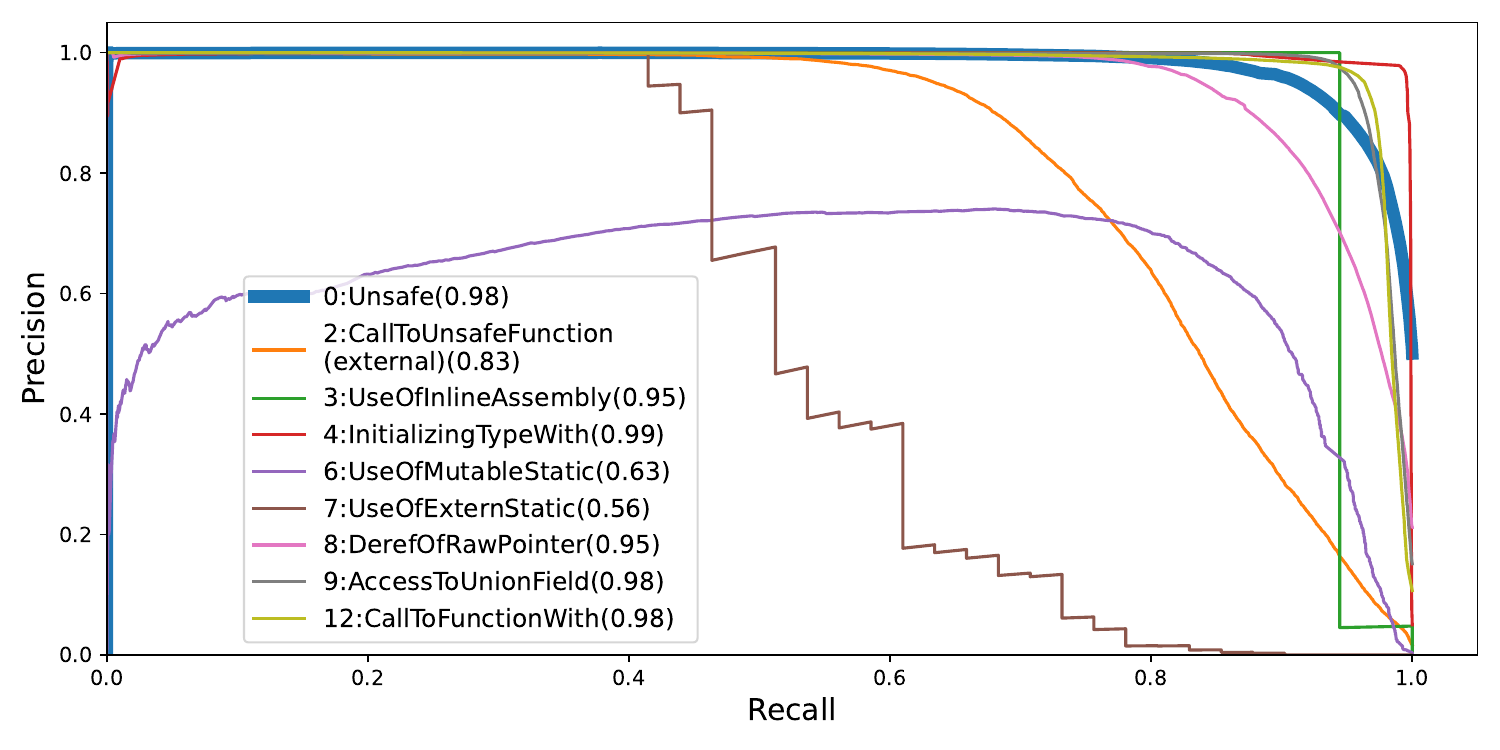}
  \caption{
    Precision-recall (AUPRC) evaluation on x64 binaries  in \rustsecunsafe{} dataset, \implname{} achieves 0.98 AUPRC score with a precision 95.21\% and recall 91.09\%, similar to the \crateunsafe{} dataset result.
  }
  \label{fig:rustsecu}
\end{figure}

To further demonstrate the classification performance of \implname{}, we evaluated \implname{} in the \rustsecunsafe{} dataset. The crates in \rustsecunsafe{} are excluded from \crateunsafe{} prior to training, so \implname{} has never seen them before. The evaluation result is shown in \autoref{fig:rustsecu}.  In general, \implname{} achieves the 0.98 AUPRC score with precision 95.21\% and recall 95.09\%, showing its ability to recover unsafe Rust operations from unknown assembly.

\subsection{RQ2: Assistance in Reverse Engineering}

The ultimate goal of \implname{} is to accelerate the bug hunting process for reverse engineers; therefore, to further demonstrate the benefit of \implname{} in RQ2, we evaluate \implname{} in the \rustsecbug{} dataset to show its ability to accelerate the bug hunting process. Then we measure the analysis speed to show its practical applications.

\subsubsection{Assisting Reverse Engineering}
The ultimate objective of \implname{} is to help reverse engineers narrow the potential search space. Therefore, to show \implname{}'s efficiency, we compare \implname{}'s guidance with following methods on the \rustsecbug{} dataset:
\sqlist
\item \textbf{Oracle}: using source code to get all unsafe operations;
\item \textbf{Random}: reverse engineers randomly pick functions;
\item \textbf{Default}: reverse engineers uses the default ordering from binary analysis tools.

\sqend

We set our target recall to be at least 80\% and the result shows \implname{} can minimize the searching space to only \finalcoverage{} and guarantees \finalrecall{} of unsafe operations inside, details are in \autoref{fig:final-result}.
Among the four methods, \implname{} performs close to the optimal case: with only 1.25x coverage overhead. Compared to randomly searching or prioritizing third-party crates, \implname{} achieves 4-6x benefits.

 \pgfplotsset{compat=1.12}
\pgfplotstableread[col sep=comma]{msan.dat}\msandata

\begin{figure}
 \resizebox{\columnwidth}{!}{
\begin{tikzpicture}[font=\footnotesize]

\definecolor{clr0}{RGB}{235, 230, 68}
\definecolor{blue1}{RGB}{87, 154, 195}
\definecolor{clr2}{RGB}{238, 171, 109}
\definecolor{clr3}{RGB}{166, 207, 159}
\definecolor{clr4}{RGB}{126, 125, 180}
\definecolor{green5}{RGB}{166, 207, 159}
\definecolor{blue6}{RGB}{103, 241, 210}
\begin{axis}[width=7cm, height=3cm,
name=myaxis,
xtick=data,
  xticklabels from table={\msandata}{subject},
  nodes near coords,
  every node near coord/.append style={font=\tiny, color=black,
  /pgf/number format/.cd, 
            fixed,               
            fixed zerofill,      
            precision=2,         
        /tikz/.cd              
  },
  extra y ticks=80,
extra y tick labels={80},
extra y tick style={
            ymajorgrids=true,
            ytick style={
                /pgfplots/major tick length=0pt,
            },
            grid style={
                red,
                dashed,
                /pgfplots/on layer=axis foreground,
            },
        },
  ybar=2pt,
  ymin=0,
    legend style={draw=none, legend columns=0},
       axis x line*=bottom,
   axis y line*=left
]
\addplot [color=blue1, fill] table [x expr=\coordindex, y={recall}] \msandata;
\addplot [color=green5, fill] table [x expr=\coordindex, y={coverage}] \msandata;
\end{axis}
\end{tikzpicture}
}

 \caption{\implname{}'s overall performance for narrowing down search space. We set the same target recall value as 80\% (blue bar) and compare the number of candidates (green bar, lower value is better) picked by each method.}
 \label{fig:final-result}
\end{figure}

\subsubsection{Performance Overhead}
\implname{}'s performance is affected by the bootstrapping process like decompiling stage and model loading stage. We sampled 200 binaries based on their size and applies \implname{} on them with single CPU core and single GPU card. It takes \implname{} around 10 hours to finish the analyses. Considering \crateunsafe{} has around 100K binary programs, it only take \implname{} for around one week to analyze all the binary programs in crate.io with 32 single CPU and GPU processes.


\subsubsection{Real World Applications}
\begin{table}[]
  \centering

  \begin{tabular}{c|c||c}
    \toprule
    category & name (LOC) & AUPRC ($\uparrow$)
    \\
    \midrule
    Web browser & \texttt{servo(11.10M)} \citep{servo} & $0.890$
    \\
    \hline
    \makecell{Ruby interpreter} &  \texttt{artichoke(1.93M)} \citep{artichoke} & $0.839$ 
    \\
    \hline
    \makecell{Python interpreter} & \texttt{rustpython(8.02M)} \citep{deno} & $0.832$ 
    \\
    \hline
    \makecell{JavaScript runtime} & \texttt{deno(9.36M)} \citep{deno} & $0.907$ 
    \\
    \bottomrule
  \end{tabular}

  \caption{
    Evaluation of \implname{}'s unsafe classification on large applications.  \implname{} achieves high AUPRC scores on all applications and even 0.907 on \texttt{deno}. 
  }
  \label{table:apps}
\end{table}

We evaluated \implname{} in real-world applications to demonstrate its capability in analyzing large binary programs that exhibit complex logic.
In particular, we choose three types of binaries:
\texttt{servo} as a web browser,
\texttt{artichoke}, \texttt{rustpython} as high-level language interpreters, Ruby and Python accordingly,
\texttt{deno} as a JavaScript and TypeScript runtime\footnote{We use servo with commit \href{https://github.com/servo/servo/commit/16da1c2721d471277c3981795d8d6000e8876cea}{16da1c2}, artichoke with commit \href{https://github.com/artichoke/artichoke/commit/4c72aba482dc53d377e5311b04babde24e46e100}{4c72aba}, RustPython with commit \href{https://github.com/RustPython/RustPython/commit/3b6db8e21a73a816bce2644abd58a5bc3d545743}{3b6db8e} and deno with commit \href{https://github.com/denoland/deno/commit/8c2f1f5a55a2a9bb9e04c12236faa341b3fd49b6}{8c2f1f5}.}, where
these binaries are not in our dataset,
while related packages might be included (\eg \texttt{smallvec} for \texttt{servo}).
We additionally count the lines of Rust code in the target repository to evaluate our model's performance. Since Rust utilizes \texttt{cargo} to manage its dependencies, we employ \texttt{cargo vendor} to download all dependencies and subsequently count the code lines, which encompass both the project's source code and all its dependencies.

\autoref{table:apps} shows the evaluation results for each application and \implname{}'s AUPRC score.
Overall, \implname{} performs well in AUPRC, achieving over 0.907 for \texttt{deno} and above 0.83 for the other applications. For language interpreters like \texttt{artichoke} and \texttt{rustpython}, \implname{}'s performance is influenced by the various system calls and low-level APIs supported by the target language. We note that all of these applications contain more than 1M lines of Rust code, and \implname{} can complete the analysis of these projects in three hours.

\subsection{RQ3: Robustness Evaluation}

By the definition of unsafe Rust in \autoref{sec:bg-memory}, it is a language-level definition that is expected to be independent of different compiler toolchains, architectures, etc. To show the robustness of \implname{}, we evaluate \implname{} across different compiler toolchains (Rustc 1.57.0, 1.67.0, and 1.75.0) and architectures(x64 and ARM).

\subsubsection{Different Compiler Toolchains}
\begin{figure}[]
  \centering
    \includegraphics[width=\columnwidth]{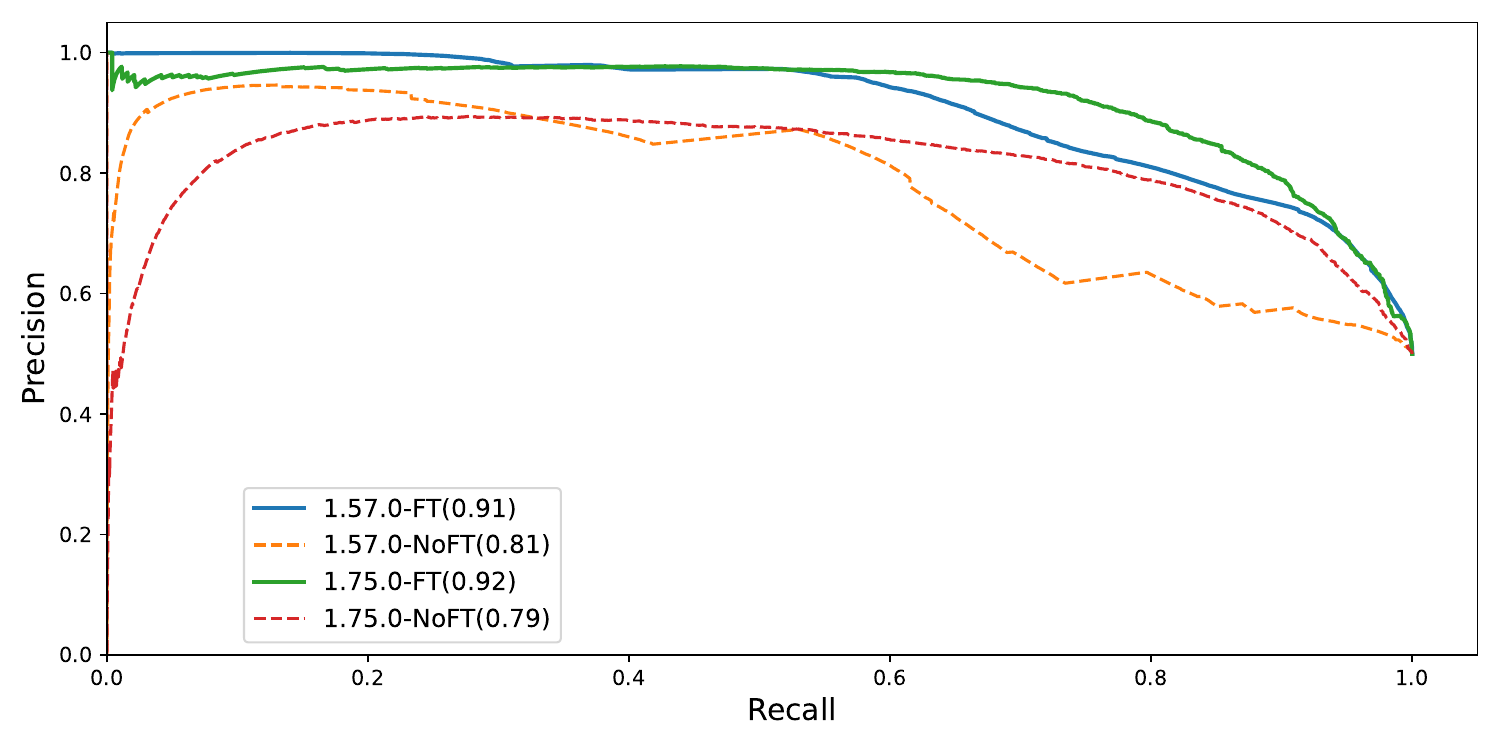}
  \caption{
    Precision-recall (AUPRC) evaluation on different compiler versions of unsafe Rust classification. With the help of fine-tuning, \implname{} improves the scores to 0.91 for 1.57.0 and 0.92 for 1.75.0.
  }
  \label{fig:unsafe-ft}
\end{figure}

We first demonstrate how fine-tuning aids \implname{} in managing various toolchains of Rust and in recovering the unsafe regions.

\smallskip\noindent\textbf{Setup.}
The \crateunsafe{} dataset is built on customized Rustc 1.67.0~\cite{rustblog167}, which uses LLVM-15 as the backend. To further demonstrate \implname{}'s robustness, we collected the \rustsecunsafe{} dataset using Rustc 1.57.0~\cite{rustblog157} with the LLVM-13 backend and Rustc 1.75.0~\cite{rustblog175} with the LLVM-17 backend, fine-tuning and evaluating \implname{}'s performance on both versions and comparing it with the 1.67.0 version. Since Rustc converts MIR into LLVM IR and leverages LLVM to optimize and generate instructions, different LLVM backend versions can produce varying outputs. We sampled only 20\% of the data from both Rustc 1.57.0 and 1.75.0, fine-tuned for two epochs in less than an hour.  

\smallskip\noindent\textbf{Result.} The comparison of the unsafe classification task is presented in \autoref{fig:unsafe-ft}. The results indicate that prior to fine-tuning, the model's score diminishes by approximately 0.10 due to the compiler toolchain changes. With the help of the fine-tuning process, \implname{} is able to capture minor changes across different compiler versions, achieving high scores among all compiler versions: 0.91 for 1.57.0 and 0.92 for 1.75.0.

\subsubsection{Different Architectures}
\begin{figure}[]
  \centering
    \includegraphics[width=\columnwidth]{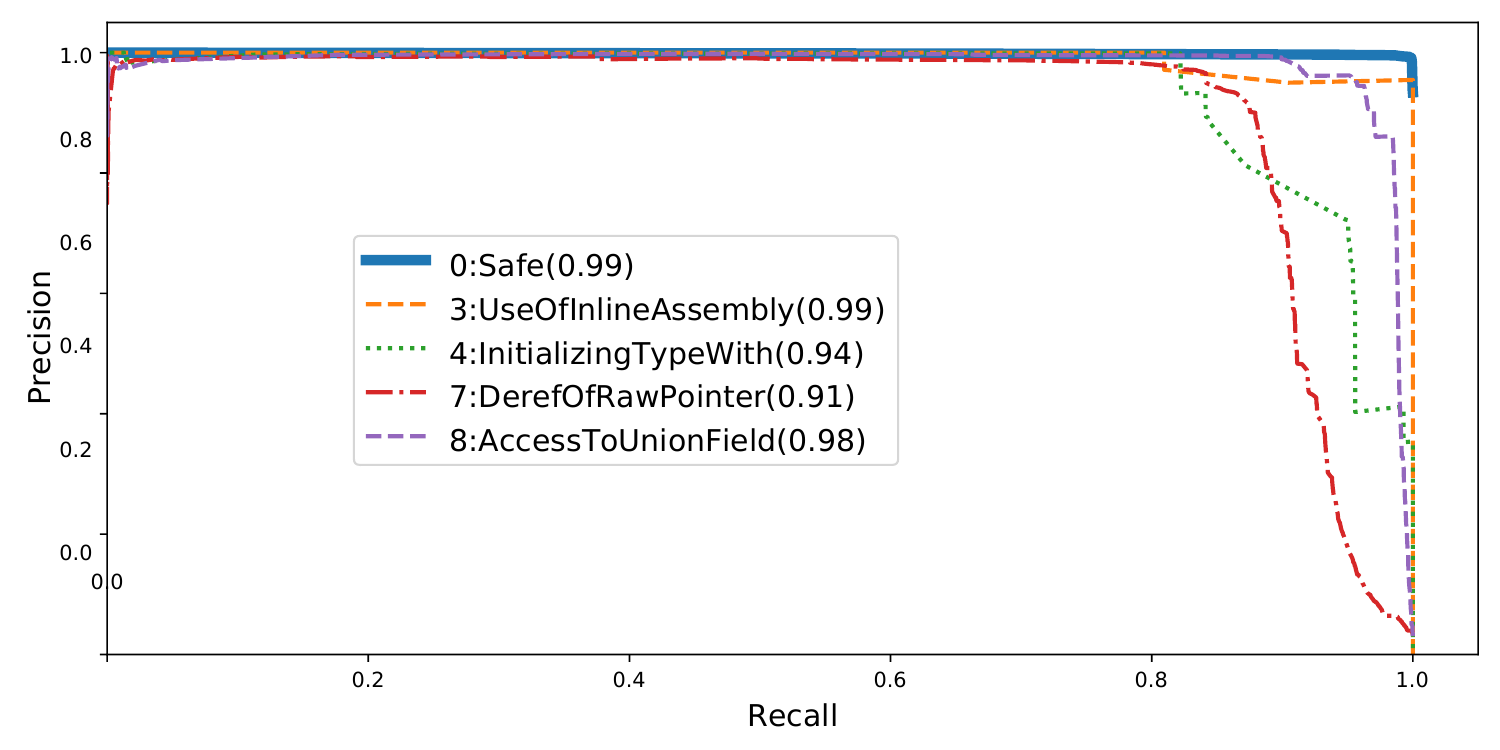}
  \caption{
    Precision-recall (AUPRC) evaluation on ARM binaries  in \crateunsafe{} dataset. \implname{} achieves high scores as x64, showing its robustness on different architectures.
  }
  \label{fig:unsafe-arm}
\end{figure}
Besides the toolchains, we further explore \implname{}'s performance under  different architectures.

\smallskip\noindent\textbf{Setup.} According to the definition of unsafe Rust, unsafe information is lost in the early stages of compilation when Rustc ports its MIR to the LLVM IR, indicating the architecture independence of unsafe betrayal. To validate this important property, we conduct a similar machine learning pipeline: dataset collection, preprocessing, model training, and evaluation on ARM architectures. 

\smallskip\noindent\textbf{Result.} The results are shown in \autoref{fig:unsafe-arm}. \implname{} demonstrate high performance on both ARM and x64 architectures, indicating the architectural independence of unsafe Rust. Some improvements observed in certain labels can be attributed to ARM's fixed-length instruction set, which offers a simpler assembly language relative to the variable-length instruction set of x64.

\subsubsection{Ablation Study}

\smallskip\noindent\textbf{Setup.} \implname{} embeds the static analysis result in the input of the model to help the model identify the inner connections. To demonstrate the effectiveness of this approach, we picked the \texttt{UseOfMutableStatic} label and retrained it without embedding the static analysis using the same dataset \crateunsafe{}. 

\begin{figure}[]
  \centering
    \includegraphics[width=0.9\columnwidth]{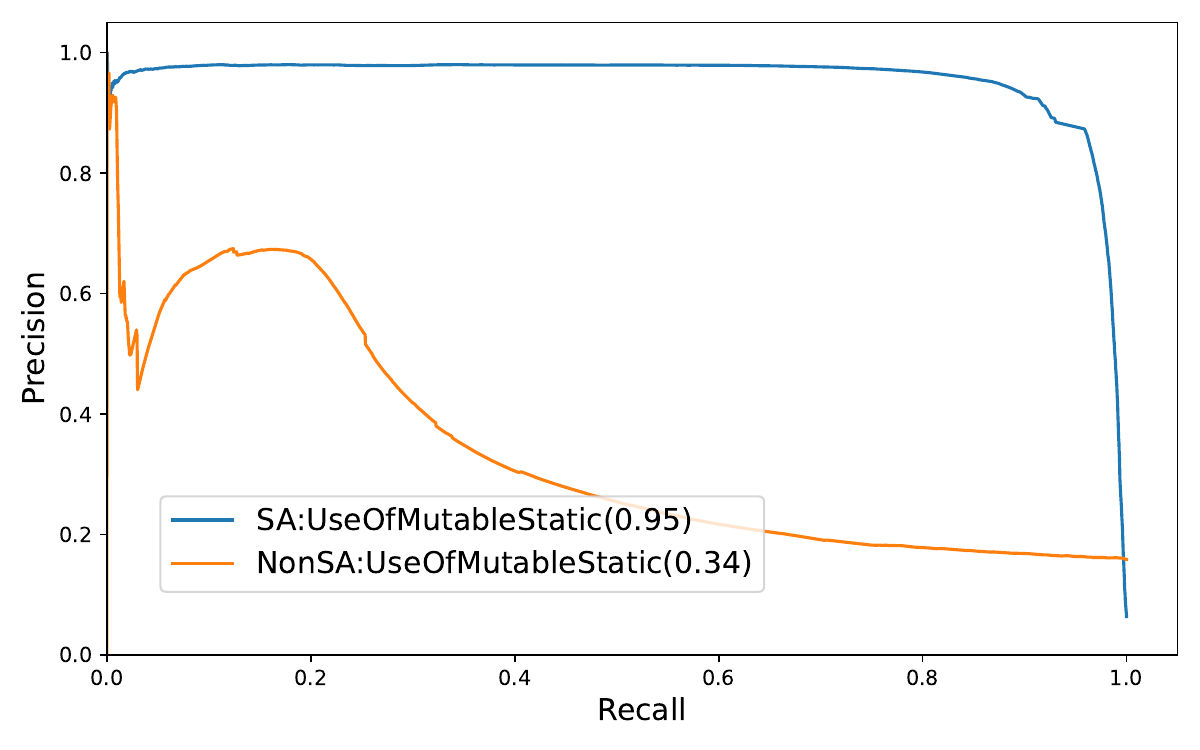}
  \caption{
    Ablation study of static analysis on x64 binaries in \crateunsafe{}{} dataset. `SA' means static analysis. The static analysis result helps \implname{} improves the performance of target label classification task by 0.61 in AUPRC.
  }
  \label{fig:sa-eval}
\end{figure}

\smallskip\noindent\textbf{Result.} The comparison is shown in \autoref{fig:sa-eval}, the static analysis helps \implname{} improve the AUPRC by 0.61, showing the benefit of embedding the static analysis result into the model input. 

\subsection{RQ4: Bug Hunting Applications}
\label{sec:eval:RQ4}

We applied \implname{} on the Angr, directed fuzzing and Android system fuzzing to assess its end-to-end effectiveness.

\subsubsection{Angr Analysis}\label{exp:angr}
\begin{table*}[]
\centering
\begin{tabular}{rr|ccc|ccc}
\toprule
\multicolumn{1}{c|}{\multirow{2}{*}{RUSTSEC}}   & \multirow{2}{*}{Name} & \multicolumn{3}{c|}{x86\_64}                                                  & \multicolumn{3}{c}{ARM}                                                 \\ \cline{3-8} 
\multicolumn{1}{c|}{}                           &                       & \multicolumn{1}{c|}{Baseline}  & \multicolumn{1}{c|}{Oracle}   & \implname{}     & \multicolumn{1}{c|}{Baseline} & \multicolumn{1}{c|}{Oracle}   & \implname{}     \\ \midrule \midrule
\multicolumn{1}{c|}{\multirow{2}{*}{2021-0015}} & search\_error         & \multicolumn{1}{c|}{7060.73}   & \multicolumn{1}{c|}{660.14}   & 1607.65  & \multicolumn{1}{c|}{29028.36} & \multicolumn{1}{c|}{3491.61}  & 1465.16  \\ \cline{2-8} 
\multicolumn{1}{c|}{}                           & excel\_to\_csv        & \multicolumn{1}{c|}{4668.81}   & \multicolumn{1}{c|}{343.36}   & 424.62   & \multicolumn{1}{c|}{2548.93}  & \multicolumn{1}{c|}{6750.60}  & 343.76   \\ \hline
\multicolumn{1}{c|}{\multirow{2}{*}{2020-0043}} & bench-server          & \multicolumn{1}{c|}{43200(TO)} & \multicolumn{1}{c|}{2908.19}  & 20565.21 & \multicolumn{1}{c|}{22305.18} & \multicolumn{1}{c|}{19151.26} & 6781.42  \\ \cline{2-8} 
\multicolumn{1}{c|}{}                           & external\_shutdown    & \multicolumn{1}{c|}{43200(TO)} & \multicolumn{1}{c|}{18805.00} & 39182.24 & \multicolumn{3}{c}{N/A}                                                 \\ \hline
\multicolumn{1}{c|}{2021-0009}                  & crosstalk             & \multicolumn{1}{c|}{26816.39}  & \multicolumn{1}{c|}{244.69}   & 5601.20  & \multicolumn{3}{c}{N/A}                                                 \\ \hline
\multicolumn{1}{c|}{2021-0088}                  & worldbank             & \multicolumn{1}{c|}{1641.82}   & \multicolumn{1}{c|}{2264.12}  & 3269.44  & \multicolumn{1}{c|}{3161.93}  & \multicolumn{1}{c|}{3725.85}  & 2972.57  \\ \hline
\multicolumn{1}{c|}{\multirow{3}{*}{2021-0092}} & extension1            & \multicolumn{1}{c|}{8672.45}   & \multicolumn{1}{c|}{263.82}   & 506.95   & \multicolumn{1}{c|}{3439.36}  & \multicolumn{1}{c|}{317.63}   & 616.76   \\ \cline{2-8} 
\multicolumn{1}{c|}{}                           & extension2            & \multicolumn{1}{c|}{10425.88}  & \multicolumn{1}{c|}{1351.28}  & 24621.30 & \multicolumn{1}{c|}{4495.87}  & \multicolumn{1}{c|}{1391.97}  & 4837.85  \\ \cline{2-8} 
\multicolumn{1}{c|}{}                           & stream                & \multicolumn{1}{c|}{6283.93}   & \multicolumn{1}{c|}{4.70}     & 423.35   & \multicolumn{1}{c|}{2385.22}  & \multicolumn{1}{c|}{887.84}   & 484.90   \\ \hline
\multicolumn{1}{c|}{\multirow{2}{*}{2021-0094}} & predefined            & \multicolumn{1}{c|}{6551.97}   & \multicolumn{1}{c|}{5050.39}  & 4598.43  & \multicolumn{1}{c|}{14804.18} & \multicolumn{1}{c|}{495.10}   & 17458.97 \\ \cline{2-8} 
\multicolumn{1}{c|}{}                           & file\_watcher         & \multicolumn{1}{c|}{27198.87}  & \multicolumn{1}{c|}{3982.78}  & 11730.89 & \multicolumn{1}{c|}{15422.56} & \multicolumn{1}{c|}{3811.30}  & 13806.29 \\ \hline
\multicolumn{1}{c|}{2021-0090}                  & texture               & \multicolumn{1}{c|}{43200(TO)} & \multicolumn{1}{c|}{8618.49}  & 4396.34  & \multicolumn{3}{c}{N/A}                                                 \\ \hline
\multicolumn{1}{c|}{\multirow{2}{*}{2021-0085}} & binjs\_dump           & \multicolumn{1}{c|}{11034.37}  & \multicolumn{1}{c|}{481.09}   & 546.64   & \multicolumn{1}{c|}{33386.22} & \multicolumn{1}{c|}{771.72}   & 589.74   \\ \cline{2-8} 
\multicolumn{1}{c|}{}                           & binjs\_decode         & \multicolumn{1}{c|}{43200(TO)} & \multicolumn{1}{c|}{3043.78}  & 1603.98  & \multicolumn{1}{c|}{3368.33}  & \multicolumn{1}{c|}{2636.86}  & 2404.86  \\ 
\midrule \midrule
\multicolumn{2}{c|}{average(seconds)}                                   & \multicolumn{1}{c|}{20225.37}  & \multicolumn{1}{c|}{3430.13}  & 8505.59  & \multicolumn{1}{c|}{12213.29} & \multicolumn{1}{c|}{3948.34}  & 4705.66  \\ 
\bottomrule
\end{tabular}
\caption{Angr analysis performance for Rust binaries in different architectures. \texttt{TO} denotes for timeout and failed to find the bug and \texttt{N/A} denotes for failed to get the buggy binary. \implname{} can save 57.95\% of time to find the same bug compared with baseline in x64 and 61.4\% for ARM binaries. Compared with the oracle baseline including source code, \implname{} is only 2.48x compared with oracle method with source code in x64 and 1.19x in ARM. }
\label{tab:angr}
\end{table*}

Angr~\citep{shoshitaishvili2016state} is a powerful binary analysis tool that finds bugs through symbolic execution. However, performing symbolic execution is time and resource-consuming, as it involves the exploration of different paths in the program. Therefore, \implname{} can be integrated as a guiding framework, enabling the prioritization of suspect functions. 

For instance, in the \texttt{predefined} binary of RUSTSEC-2021-0094, there are 1,434 functions in total, and under the default ordering the actual buggy function is ranked 783. \implname{} reorders functions according to their estimated unsafe probabilities, elevating the buggy function to rank 140 and accelerating the bug localization process.

\PP{Setup}We collected the benchmark from the RustSec dataset, manually filtered out bugs that were not shown in the binary programs, and successfully built 14 vulnerable binaries. We set up our Angr analyzes based on QueryX~\cite{han2023queryx}'s memory safety analysis script, identifying heap/stack overflows, use-after-free, and out of bounds access errors, and added detection of uninitialized memory access. We established targets for each function in the binaries with a total limitation of 12 hours per binary and 5 minutes per function. We compared our approach with the baseline, which provided no guidance, and the unsafe oracle from the source code; the results are shown in \autoref{tab:angr}.

\PP{Result}\implname{} is close to the unsafe oracle result, with only 1.48x overhead to reach the oracle case with additional source code information. Compared to the baseline approach, \implname{} saves 57.95\% of time to find the same bugs, largely saving time and resources during the vulnerability hunting process.We also performed the same evaluation on the ARM binaries, and \implname{} saves 61.4\% on ARM architectures. 

\subsubsection{Fuzzing Analysis}
\label{exp:fuzzing}

\begin{table}[]

\resizebox{.99\columnwidth}{!}{%
\begin{tabular}{c|r|r|r|r}
\toprule
\textbf{Issue} & \textbf{Target}       & \textbf{Baseline} & \textbf{\implname{} Guided}   & \textbf{Ratio}   \\ 
\midrule
\midrule
\href{https://github.com/3Hren/msgpack-rust/issues/151}{\faGithubSquare} & rmpv          & 7.80     & 6.20     & 79.4\%  \\ \hline
\href{https://github.com/alex/rust-asn1/issues/32}{\faGithubSquare}& asn           & 86.78    & 35.04    & 40.4\%  \\ \hline
\href{https://github.com/image-rs/image-tiff/issues/28}{\faGithubSquare}& tiff          & 158.44   & 132.58   & 83.7\%  \\ \hline
\href{https://github.com/serde-rs/serde/issues/82}{\faGithubSquare}& serde         & 5081.94  & 4931.55  & 97.0\%  \\ \hline
\href{https://github.com/dtolnay/proc-macro2/issues/55}{\faGithubSquare}& proc\_macro2  & 10360.74 & 8048.14  & 77.7\%  \\ \hline
\href{https://github.com/boa-dev/boa/issues/771}{\faGithubSquare}& boa           & 27791.16 & 26474.50 & 95.3\%  \\ \hline
\href{https://github.com/gimli-rs/cpp_demangle/pull/41}{\faGithubSquare}& cpp\_demangle & 31128.52 & 30317.32 & 97.4\%  \\ 
\midrule
\midrule
&average           & 10659.34 & 9959.88  & 78.7\% \\ 
\bottomrule
\end{tabular}%
}
\caption{\implname{} provides guided targets for AFLGO. On average, \implname{} can save 21.26\% of the fuzzing time to find the crash.}
\label{tab:fuzzing}
\end{table}

Directed fuzzers are emerging dynamic analysis tools that leverage control flow graphs and static analysis to compute the instruction's distance to the target function and prioritize the corpus that reaches closer locations.

\PP{Setup}We apply the result of \implname{} to a directed fuzzer as its target function to show the direction of \implname{} for dynamic analysis.
 Specifically, we use AFLGO \cite{bohme2017directed} as our directed fuzzer, and we use the trophy cases found by \texttt{cargo fuzz} as our benchmark. Among the 7 reproducible bugs with their harnesses, we first launch the undirected AFL, then apply \implname{} to these binaries and launch the AFLGO with \implname{}'s output as the target. We count the wall time of the fuzzers that encounter the first crash as a result and repeat the fuzzing process three times to obtain the mean value. 

\PP{Result}The result is shown in \autoref{tab:fuzzing}; on average, with the help of \implname{}, AFLGO can save 21.26\% of the time to find the same bug without guidance.  

\subsubsection{Android Rust Library Fuzzing}To evaluate \implname{}'s end-to-end effectiveness, we applied \implname{} to the Android system and provided guidance for fuzzing Rust libraries.

\PP{Setup}As of Android 16.0.0, there are 93 Rust crates serving as fundamental OS libraries, containing 1,534 function APIs. Testing all of the functions requires a lot of engineering effort. We consider the following steps:
given a crate library in \texttt{.dylib or .so}, (1) we feed all the functions into \implname{} and get a prioritized target list; manually develop the fuzzing harnesses for each function,  
and
(2) conduct black box fuzzing with AFL++ in Frida mode~\cite{valsamarasfuzzing2024} for the top 50 targets. For each target, we run for a maximum of 24 hours. 

\PP{Result}We identified and confirmed five different bugs: two stemming from character boundary issues, one concerning an out-of-bounds access, one related to an unexpected unwrap in the library, and one resulting in a panic abort\footnote{The bugs are available at: \url{https://issuetracker.google.com/issues/399131919}}. We reported all the PoC code with corresponding inputs to Google, and all the bugs were confirmed and patched by the maintainers. Compared with the default approach of fuzzing each function individually, \implname{} saves 74.6\% of time to find these bugs.

\section{Limitation}
\vspace{-0.5ex}
\label{sec:dis}

\implname{}'s limitation can be categorized into two parts:
the limitation inherited from the toolsets and methodologies used by \implname{},
and the limitation related to our implementation.

\PP{Inherited Limitations}\implname{} relies on Rustc to generate the corresponding dataset for training purposes. For unsafe operations that Rustc cannot detect, \implname{} cannot identify them as well.
Second, \implname{} leverages machine learning to recover unsafe Rust, the loss during the training process cannot be recovered by \implname{}.
\implname{} experiences instances of false positives and false negatives compared to the oracle method. 

\PP{Implementation Limitations}\implname{}'s implementation limitations are mainly from the data collection and model training steps. Currently \implname{} leverages the specific Rustc to automatically generate the \crateunsafe{} dataset. Because of the compiler differences and architecture requirements, \crateunsafe{} dataset may miss several crates and the unsafe functions are missed by \implname{} as well.
Furthermore, constrained by hardware capacity, \implname{} utilizes only 10M out of a 903M dataset for training. Finally, as a prototype, \implname{} uses RoBERTa~\cite{liu2019roberta}, which is a classic BERT model specialized for classification tasks.
Finally, due to the context window limitation, \implname{} may be inaccurate because of different context splits.

\section{Related work}
\vspace{-0.5ex}
\para{Unsafe Rust and its usage.}
Although software engineers try to avoid \texttt{unsafe} regions in their program to avoid potential memory safety problems \cite{fulton2021benefits, cui2024unsafe}, many Rust crates are using "unsafe" more frequently and lead to implicit usage and wide spread of unsafe blocks in Rust binaries \cite{evans2020rust}.
Also, studies \cite{qin2020understanding, astrauskas2020programmers} show that these unsafe usages of \texttt{} are often for good or unavoidable reasons, which are not easy to remove.
While \texttt{unsafe} catches the attention of programmers on memory safety, it can be also used by reverse engineers to find the weakness of the given Rust binaries. 
To our knowledge, \implname{} is the first tool to recover the \texttt{unsafe} regions from raw binary instructions. 

\para{Binary bug hunting.}
Approaches to find bugs or vulnerabilities have been proposed through binary analyses \citep{shoshitaishvili2016state,cova2006static,cheng2018dtaint,sun2020hybrid,david2018firmup,wang2019oo7,wu2019kepler,wu2018fuze}. 
Angr \citep{shoshitaishvili2016state} is a powerful framework that combines static and dynamic analyses to automatically find general vulnerabilities in binary executable. 
In contrast, other tools are designed to find specific bugs in binaries;
oo7 \citep{wang2019oo7} is designed for \emph{spectre attacks}, KEPLER \citep{wu2019kepler} is targeted for \emph{control-flow hijacking}, and DTaint \citep{cheng2018dtaint} aims to detect \emph{taint-style} vulnerabilities.
Furthermore, machine learning has been applied to program analyses for bug finding
\citep{padmanabhuni2015buffer,li2018vuldeepecker,pradel2018deepbugs}.
VulDeePecker is a system that leverages deep learning to automatically detect bugs inside programs. However, \citep{zhao2018buffer} also proposed several open questions that may interfere with the performance of applying machine learning to the search for bugs.
%

\para{General binary analysis.}
Beyond bug hunting, general binary analysis is an essential task in computer security. Some of this research (\eg function boundary detection) can be regarded as a basis of \implname{}. While Nova~\cite{jiang2023nova} utilizes hierarchical attention and contrastive learning for general semantic representation, \implname{} is a specialized classifier designed for the security-critical task of identifying \textbf{direct unsafe operations}. 
In \emph{binary-binary code matching}, a binary function or an entire program is represented in a vector to retrieve functions or programs similar to a binary target given \citep{ding2019asm2vec,yu2020order,peng2021how}. The core of known approaches is learning binary code representation to summarize the instructions of a function or a program into a vector based on recurrent neural networks \citep{hochreiter1997long}, graph neural networks \citep{scarselli2008graph}, or transformer-based models \citep{vaswani2017attention}. 
Similarly, \emph{binary-source code matching} finds similar binary or source code given source or binary code \citep{yu2020codecmr}. 
\emph{Function prototype inference} is predicting the type of function (\eg the number of arguments and the type of argument) given instructions of a function \citep{chua2017functype}. 
\emph{Function boundary detection} enumerates the list of the start and end of a function in a binary \citep{idapro,shin2015recognizing,koo:func-identification,di2017rev}, and \emph{malware classification} classifies each binary regardless of whether it embeds malware \citep{raff2018malware}.

\section{Conclusion}
\vspace{-0.5ex}
In this paper,
we present \implname{},
the first tool to leverage machine learning and static analysis
to identify unsafe regions in Rust binaries.
%
Our evaluation shows that \implname{} can identify unsafe regions from Rust binaries with precision 93.84\% and recall 91.75\%.
\implname{} can accelerate the vulnerability finding process in symbolic execution (static) and directed fuzzing (dynamic) by 57.95\% and 21.26\% respectively. By applying \implname{} to blackbox fuzzing in Android, we successfully identified 5 unknown bugs in the Rust libraries and all the bugs were confirmed by the developers.

\section{Acknowledgment}
\label{s:ack}
\vspace{-0.5ex}
We thank our shepherd Hui Xu and the anonymous reviewers for their valuable feedback and suggestions. We also thank Mingyu Guan and Zhuoran Yu for their assistance with model training.
This research was supported, in part, by the 
ONR under grant N00014-23-1-2095,
IITP grant (No. RS-2024-00509258 and No. RS-2024-00469482),
NRF grant (RS-2025-00560062),
and gifts from Facebook, Mozilla, Intel, VMware and Google.

\clearpage
\bibliographystyle{plain}
\bibliography{twsml}

\end{document}